\documentclass[11pt,a4paper]{article}
\pdfoutput=1
\usepackage{jcappub}
\allowdisplaybreaks
\usepackage{graphicx,psfrag}
\usepackage{bm}
\usepackage{mathbbol,verbatim}
\usepackage{slashed}
\usepackage{graphics}
\usepackage{color,ulem}

\allowdisplaybreaks

\definecolor{greeen}{rgb}{0.03,0.84,0.13}
\definecolor{test}{rgb}{0.03,0.74,0.33}
\definecolor{viol}{rgb}{0.44,0,0.94}
\definecolor{or}{rgb}{0.95,0.65,0}

\newcommand{\TianQin}{TianQin~\cite{Luo:2015ght}}
\newcommand{\Taiji}{Taiji~\cite{Guo:2018npi}}
\newcommand{\LISA}{LISA~\cite{Audley:2017drz, Cornish:2018dyw}}
\newcommand{\ALIA}{ALIA~\cite{Gong:2014mca}}
\newcommand{\MAGIS}{MAGIS~\cite{Coleman:2018ozp}}
\newcommand{\DECIGO}{DECIGO~\cite{Musha:2017usi}}
\newcommand{\BBO}{BBO~\cite{Corbin:2005ny}}
\newcommand{\aLIGO}{aLIGO~\cite{LIGOScientific:2019vkc}}
\newcommand{\aLIGOp}{aLIGO+~\cite{aLIGO+}}
\newcommand{\ET}{ET~\cite{Punturo:2010zz}}
\newcommand{\CE}{CE~\cite{Evans:2016mbw}}

\begin{document}

\title{Gravitational Waves from First-Order Phase Transition in a Simple Axion-Like Particle Model}

\author[a]{P. S. Bhupal Dev,}
\author[a]{Francesc Ferrer,}
\author[a]{Yiyang Zhang,}
\author[a,b]{Yongchao Zhang}
\affiliation[a]{Department of Physics and McDonnell Center for the Space Sciences,  Washington University, St. Louis, MO 63130, USA}
\affiliation[b]{Center for High Energy Physics, Peking University, Beijing 100871, China}

\abstract{We consider a gauge-singlet complex scalar field $\Phi$ with a global $U(1)$ symmetry that is spontaneously broken at some high energy scale $f_a$. As a result, the angular part of the $\Phi$-field becomes an axion-like particle (ALP). We show that if the $\Phi$-field has a non-zero coupling $\kappa$ to the Standard Model Higgs boson, there exists a certain region in the $\left(f_a, \kappa\right)$ parameter space where the global
$U(1)$ symmetry-breaking induces a strongly first order phase transition,
thereby producing stochastic gravitational waves that are potentially
observable in current and future gravitational-wave detectors. In particular,
we find that future gravitational-wave experiments such as TianQin, BBO and
Cosmic Explorer could probe a broad range of the energy scale
$10^3 \, {\rm GeV} \lesssim f_a \lesssim 10^{8} \, {\rm GeV}$, independent of
the ALP mass. Since all the ALP couplings to the Standard Model particles are
proportional to inverse powers of the energy scale $f_a$ (up to
model-dependent ${\cal O}(1)$ coefficients), the gravitational-wave
detection prospects are largely complementary to the current laboratory,
astrophysical and cosmological probes of the ALP scenarios.
}

\maketitle

\section{Introduction}

Axion-like particles (ALPs) are light gauge-singlet pseudoscalar bosons that couple weakly to the Standard Model (SM) and generically appear as the pseudo-Nambu-Goldstone boson (pNGB) in theories with a spontaneously broken global $U(1)$ symmetry. ALPs could solve some of the open questions in the SM, such as the strong CP problem via the Peccei-Quinn mechanism~\cite{Peccei:1977hh} and the hierarchy problem via the relaxion mechanism~\cite{Graham:2015cka}. They could also play an important cosmological role in inflation~\cite{Freese:1990rb, Adams:1992bn, Daido:2017wwb}, dark matter (DM)~\cite{Preskill:1982cy, Abbott:1982af, Dine:1982ah}, dark energy~\cite{Jain:2004gi, Kim:2009cp, Kim:2013jka, Lloyd-Stubbs:2018ouj}, and baryogenesis~\cite{Daido:2015gqa, DeSimone:2016bok}. Furthermore, there are recent proposals involving axions to simultaneously address several open issues of the SM in one stroke~\cite{Salvio:2015cja, Ballesteros:2016euj, Ballesteros:2016xej, Ema:2016ops, Salvio:2018rv, Gupta:2019ueh}.

A common characteristic among ALPs is that their coupling to SM particles is suppressed by inverse powers of the $U(1)$ symmetry breaking energy scale $f_a$. This energy scale can be identified as the vacuum expectation value (VEV) of a SM-singlet complex scalar field $\Phi$, i.e. $\langle \Phi\rangle=f_a/\sqrt{2}$, which is assumed to be much larger than the electroweak scale $v_{\rm ew}\simeq 246.2$ GeV to evade current experimental limits~\cite{Jaeckel:2010ni, Irastorza:2018dyq}. The ALP field $a$ then arises as the massless excitation of the angular part of the $\Phi$-field:
\begin{align}
\Phi(x) \ = \ \frac{1}{\sqrt 2}\left[f_a+\phi(x)\right]e^{ia(x)/f_a} \, .
\label{eq:phix}
\end{align}
The particle excitation of the modulus $\phi$ of the $\Phi$-field gets a large mass $m_\phi\sim f_a$, while the angular part $a$ becomes a pNGB that acquires a much smaller mass $m_a$ from explicit low energy $U(1)$-breaking effects. Thus, for the low-energy phenomenology of ALPs, the modulus part $\phi$ can be safely integrated out, and the only experimentally relevant parameters are $m_a$ and $f_a$.

In this paper, we show that the dynamics of the modulus $\phi$-field around the $f_a$ scale can provide complementary constraints on the ALP scenario. In particular, if the parent $\Phi$-field has a non-zero coupling to the SM Higgs doublet $H$, the $U(1)$ symmetry breaking at the $f_a$-scale could induce a strongly first-order phase transition (FOPT)~\cite{Barger:2008jx}, giving rise to stochastic gravitational waves (GWs) that are potentially observable in current and future GW detectors (see e.g.~\cite{Weir:2017wfa, Mazumdar:2018dfl} for a review on GWs from a FOPT). We find that GW signals of strength up to $h^2 \Omega_{\rm GW} \sim 10^{-12}$ could be generated,
where $\Omega_{\rm GW}$ is the fraction of the total energy density of the universe in the form of GWs today and $h=0.674\pm 0.005$ is the current value of the Hubble parameter in units of 100 km s$^{-1}$ Mpc$^{-1}$~\cite{Aghanim:2018eyx}. Future GW observatories like
\TianQin, \Taiji, LISA~\cite{Audley:2017drz, Cornish:2018dyw}, ALIA~\cite{Gong:2014mca}, MAGIS~\cite{Coleman:2018ozp}, DECIGO~\cite{Musha:2017usi}, BBO~\cite{Corbin:2005ny}, Cosmic Explorer (CE)~\cite{Evans:2016mbw} and Einstein Telescope (ET)~\cite{Punturo:2010zz} can probe a broad range $10^3 \text{ GeV} \lesssim f_a \lesssim 10^8 \text{ GeV}$, {\it independent} of the ALP mass. It turns out that the \aLIGO~ and \aLIGOp~ can not probe any of the allowed parameter space for the benchmark configurations considered in this paper (see Fig.~\ref{fig:signal-vary}).

The heavy modulus $\phi$ decouples at low energies, and we are left with
the ALP $a$, which has only derivative couplings to the SM particles. These
are generated via effective higher-dimensional operators~\cite{Brivio:2017ije}, and are proportional to inverse powers of $f_a$, up to model-dependent ${\cal O}(1)$ coefficients.
The effective ALP couplings to photons, electrons and nucleons are strongly
constrained by a number of laboratory, astrophysical and cosmological
observables~\cite{Irastorza:2018dyq}. However, current and future
low-energy constraints depend on the ALP mass $m_a$, while we find that the GW prospects in the $(m_a,f_a)$ plane are largely
complementary. For instance, if a stochastic GW signal was found with the frequency
dependence predicted by the FOPT\footnote{The frequency dependence in this scenario is generically
different and can be distinguished from other stochastic GW sources, like inflation~\cite{Turner:1996ck, Easther:2006gt, Easther:2006vd, Senatore:2011sp,
Guzzetti:2016mkm, Bartolo:2016ami} or unresolved binary black hole mergers~\cite{TheLIGOScientific:2016wyq}.}, this would point to a limited range of
$f_a$ in a given ALP model, which might lead to a {\it positive} signal in some
of the future laboratory and/or astrophysical searches of ALPs. On the other hand, if we fix the ALP mass $m_a$, then current ALP constraints
exclude certain ranges of $f_a$. If a GW signal is found in the
frequency range corresponding to the excluded range of $f_a$, then the
underlying simple ALP model has to be extended to account for the GW signal.

The rest of the paper is organized as follows: in Section~\ref{sec:model} we provide the details of the ALP model, and compute the one-loop effective
scalar potential at both zero and finite temperatures. In Section~\ref{sec:GW} we calculate the GW emission from a strong FOPT at
the scale $f_a$, including bubble collision, sound wave (SW) and magnetohydrodynamic (MHD) turbulence contributions. The complementary reach of laboratory, astrophysical and cosmological observations is presented in Section~\ref{sec:complementarity}. The constraints from future precision Higgs data on the ALP model are discussed in
Section~\ref{sec:collider}. We summarize and conclude in Section~\ref{sec:conclusion}. The method used to obtain the power-law integrated sensitivity curves for future GW experiments is described in Appendix~\ref{app:A}.

\section{Scalar Potential in the ALP model}
\label{sec:model}

\subsection{Tree-level potential}
\label{sec:model:temperature0}
The coupling between the SM Higgs doublet $H$ and the complex field $\Phi$
is described by the tree-level potential (see also Refs.~\cite{Barger:2008jx, Chiang:2017nmu})
\begin{equation}
\label{eqn:tree-level-V}
{\cal V}_{0}\ = \
-\mu^{2}\vert H\vert^{2} + \lambda\vert H\vert^{4}
+ \kappa\vert\Phi\vert^{2}\vert H\vert^{2}
+ \lambda_{a} \left(\vert\Phi\vert^{2}-\frac{1}{2} f_{a}^{2} \right)^{2} \,.
\end{equation}
The SM Higgs doublet can be parameterized as $H=\left(G_{+},\left(h+iG_{0}\right)/\sqrt{2}\right)$, with $h$ the SM Higgs and $G_0$, $G_+$ the Goldstone bosons that become the longitudinal components of the $Z$ and $W^+$ bosons, respectively. The complex field $\Phi$ can be expressed in the form given by Eq.~\eqref{eq:phix}. At an energy scale around $f_a$, a  phase transition (PT) occurs which breaks the global $U(1)$ symmetry. The field $\Phi$ gets a VEV $\langle \Phi \rangle = f_a/\sqrt{2}$, and the associated pNGB $a$ is identified as the physical ALP. Depending on the parameters $f_a$, $\kappa$ and $\lambda_a$, the PT may be strongly first order, in which case it would generate a spectrum of GWs that could be detected in current or future experiments~\cite{Dev:2016feu}, as detailed in Section~\ref{sec:GW}. We note that at this stage the ALP $a$ is neither involved in the scalar potential given by Eq.~(\ref{eqn:tree-level-V}), nor in the GW emission from the high-energy scale PT. The low-energy effective couplings of $a$ to SM particles will be discussed in Section~\ref{sec:axion:couplings}.

In terms of the real scalar field components, the tree-level potential in
Eq.~(\ref{eqn:tree-level-V}) can be re-written as:
\begin{eqnarray}
\nonumber
\mathcal{V}_{0}
& \ = \ & \frac{\lambda_{a}}{4}\left(\phi^{2}-f_{a}^{2}\right)^{2}
+\left[\frac{\kappa}{2}\phi^{2}-\mu^{2}\right]
\left(\frac{1}{2}h^{2}+\frac{1}{2}G_{0}^{2} + G_{+}G_{-} \right) \\
&&
+\lambda\left[\frac{1}{2} h^{2}+ \frac12 G_{0}^{2}  + G_{+}G_{-}\right]^{2}.
\end{eqnarray}
Setting the field values of the Goldstone modes to zero, we have
\begin{equation}
\mathcal{V}_0 (\phi,h) \ = \
\frac{\lambda_{a}}{4}\left(\phi^{2}-f_{a}^{2}\right)^{2}
+\frac{\kappa}{4} \phi^2 h^2 -\frac{\mu^2}{2} h^2
+ \frac{\lambda}{4}h^4.
	\label{eq:tree-level-g0}
\end{equation}
Here $f_a$, $\mu$, $\kappa$ and $\lambda_a$ are taken as free parameters.
By examining the tree-level potential in Eq.~(\ref{eq:tree-level-g0})
we have found that a FOPT occurs along
the $\phi$ direction, while maintaining the VEV of $h$ equal to zero at the
same time, if the parameters satisfy the inequality:
\begin{equation}
\label{eqn:choice_ineq}
\mu^2 \ \leq \ \frac{2\lambda \lambda_a}{\kappa} f_a^2 \, .
\end{equation}
While this is not the only possible choice that results in a FOPT along the $\phi$ direction, for our present purpose, we will
set the $\mu$-parameter to the value that saturates the inequality above, i.e.,
\begin{equation}
\label{eqn:choice}
\mu^2 \ = \ \frac{2\lambda \lambda_a}{\kappa} f_a^2 \, .
\end{equation}
We have verified numerically that, with this choice of $\mu^2$, the VEV of $h$ remains zero up to one-loop level during the phase transition, if there is one. Therefore, in the following analysis, we will ignore the dependence of the effective potential on the $h$-field, and consider $\phi$ as the only dynamical field.

It should be noted that with the specific choice in Eq.~\eqref{eqn:choice}, the scalar $\phi$ will contribute radiatively to the SM Higgs mass, making the
latter unacceptably large at the electroweak scale. Nevertheless, this contribution can be cancelled out, e.g. by introducing vector-like fermions at the $\mu$-scale that keep the SM Higgs boson mass at the observed 125 GeV~\cite{Graham:2009gy}. In the parameter space of interest in this paper,
$\lambda_a \simeq 10^{-3}$ and $\kappa \simeq 1$ (see Section~\ref{sec:GW}),
we have $\mu \sim 10^{-2} f_a$. Therefore, the effect of these extra
vector-like fermions on the renormalization group (RG) running of the SM couplings, as required for the calculation of the critical temperature
(see Section~\ref{subsec:critical-temp}), is expected to be small and will not be considered here. Since we are
focusing on a generic ALP scenario below the $f_a$ scale, we will defer a detailed study of ultraviolet-completion involving additional heavy fields to a future work.

\subsection{Effective finite-temperature potential}

At finite temperature $T \neq 0$, the effective one-loop potential of the scalar fields is~\cite{Dolan:1973qd, Arnold:1992rz, Quiros:1999jp, Curtin:2016urg}:
\begin{equation}
\label{eqn:potential:ft}
{\cal V}(\phi,T) \ = \ {\cal V}_0(\phi) + {\cal V}_{\rm CW}(\phi) + {\cal V}_{T}(\phi,T) \,,
\end{equation}
where $\mathcal{V}_{\rm CW}$ is the Coleman-Weinberg (CW) potential~\cite{Coleman:1973jx} that contains all the
one-loop corrections at zero temperature with vanishing external momenta,
and $\mathcal{V}_{T}$ describes the finite-temperature corrections.
Working in the Landau gauge to avoid ghost-compensating terms, the CW potential reads:
\begin{equation}
\label{CW-pot}
{\cal V}_{\rm CW}\left(\phi \right) \ = \
\sum_{i}(-1)^F n_{i}\frac{m_{i}^{4}\left(\phi \right)}{64\pi^{2}}
\left[\log\frac{m_{i}^{2}\left(\phi\right)}{\Lambda^{2}}-C_{i}\right] \,.
\end{equation}
The sum runs over all the particles that couple to the $\phi$ field (notice
that massless particles do not contribute).
In Eq.~(\ref{CW-pot}), $F = 1$ for fermions and $0$ for bosons;
$n_i$ is the number of degrees of freedom of each particle;
$C_i = 3/2$ for scalars and fermions, and $5/6$ for gauge bosons;
and $\Lambda$ is the renormalization scale, which will be set to $f_a$
throughout this paper.

The finite-temperature corrections are given by:
\begin{equation}
\label{finite-T-pot}
{\cal V}_{T}\left(\phi,T\right) \ = \
\sum_{i}\left(-1\right)^{F} n_{i}\frac{T^{4}}{2\pi^{2}}
J_{B/F}\left(\frac{m_{i}^{2}\left(\phi\right)}{T^{2}}\right) \,,
\end{equation}
where the thermal functions are:
\begin{equation}
\label{thermal-func}
J_{B/F}\left(y^{2}\right) \ = \
\int_{0}^{\infty} {\rm d}x \ x^{2}\log\left[1\mp\exp\left(-\sqrt{x^{2}+y^{2}}\right)\right].
\end{equation}
Here, the minus sign ``$-$'' is for bosons and the positive sign ``$+$'' for
fermions. We also need to include the resummed daisy corrections,
that add a temperature-dependent term $\Pi_i(T)$ to
the field-dependent mass $m_i^2$~\cite{Curtin:2016urg}. To leading order,
we have in the ALP model:
\begin{eqnarray}
\label{thermal-mass-h}
\Pi_{h}\left(T\right)\ = \ \Pi_{G_{0,\pm}}\left(T\right)
& \ = \ &
\left[\frac{3}{16}g_{2}^{2}+\frac{1}{16}g_{1}^{2}+\frac{\kappa}{12}+\frac{\lambda}{2}+\frac{y_{t}^{2}}{4}\right] T^{2}, \\
\label{thermal-mass-phi}
\Pi_{\phi}\left(T\right)
& \ = \ &
\left(\frac{\kappa}{6}+\frac{\lambda_{a}}{3}\right)T^{2} \, .
\end{eqnarray}
Effectively, the mass
terms $m_i^2$ in Eqs.~(\ref{CW-pot}) and (\ref{finite-T-pot}) get replaced
by $m_i^2 + \Pi_i(T)$.

The effective potential in Eq.~(\ref{eqn:potential:ft}) could become complex
due to $m_i^2$ being negative. This is related to particle decay and does not
affect the computation of the dynamics of PT (see
e.g.~Refs.~\cite{Iliopoulos:1974ur, Weinberg:1987vp, Delaunay:2007wb} for more details). In the numerical calculations in Section~\ref{sec:GW} we will always take
the real part of the effective potential.

\label{sec:GW}

\section{First-order phase transition}
\label{sec:GW}
\begin{figure}[!t]
  \begin{centering}
  \includegraphics[height=0.25\textheight]{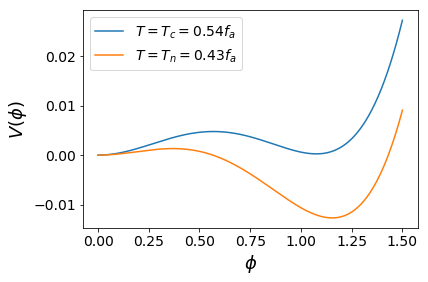}
  \par\end{centering}
  \caption{\label{fig:pot-FOPT} An example of the effective potential with a potential barrier when $T=T_c$ (see Section~\ref{subsec:critical-temp}) and $T=T_n$ (see Section~\ref{subsec:rate-nucleation}). Here $f_a = 10^6$ GeV, $\kappa=3.8$ and $\lambda_a=1.15\times 10^{-3}$.}
\end{figure}

We fix the decay constant $f_a$ at a specific value and scan
in $(\kappa, \lambda_a)$ to find the region in parameter space where a FOPT can
take place. In particular, we evaluate the effective potential for given values of $(\kappa, \lambda_a)$ to look for regions where there is a valid critical temperature $T_c$. Here, $T_c$ is defined as the temperature at which the two local minima of the effective potential are degenerate.\footnote{In our analysis, we have not considered the potential gauge-invariance~\cite{Patel:2011th} and RG improvement~\cite{Chiang:2017nmu}  effects on the calculation of $T_c$ and the resultant GW prospects, which are beyond the scope of this paper.}
We use the package {\tt CosmoTransitions}~\cite{Wainwright:2011kj} for the
numerical work and the results are given in the following subsections. We show one example of the effective potential varying temperature $T$ in Fig.~\ref{fig:pot-FOPT}. In this case, a valid $T_c$ can be found, and tunneling can happen for $T < T_c$ to give rise to a FOPT.

\subsection{Bounce solution}

The decay rate of the false vacuum is~\cite{Coleman:1977py, Linde:1980tt, Linde:1981zj}:
\begin{eqnarray}
\label{eqn:decay}
\Gamma\left(T\right) & \ \simeq \ &
\max \left[T^{4}\left(\frac{S_{3}}{2\pi T}\right)^{3/2}\exp\left(-S_{3}/T\right), \;
\left(\frac{S_{4}}{2\pi R_{0}^{2}}\right)^{2}\exp\left(-S_{4}\right) \right] \,,
\end{eqnarray}
where the first term corresponds to thermally induced decays and the second term is the quantum-tunneling rate. In Eq.~(\ref{eqn:decay}) $S_3$ and $S_4$ are the three- and four-dimensional Euclidean
actions for the $O(3)$ and $O(4)$-symmetric tunnelling (``bounce'') solutions
respectively, and $R_0$ is the size of the bubble. For quantum tunneling,
\begin{equation}
S_{4} \ = \ \int {\rm d}^{4}x
	\left[\frac{1}{2}\left(\frac{{\rm d}\phi}{{\rm d} t}\right)^{2}
+\frac{1}{2}\left(\nabla\phi\right)^{2} + {\cal V} \left(\phi,T\right)\right] \,,
\label{action-s4}
\end{equation}
where $\phi(r)$ is the solution of the $O(4)$-symmetric instanton (with $r=\sqrt{\pmb{r}^2+t^2}$):
\begin{equation}
\frac{{\rm d}^{2}\phi}{{\rm d}r^{2}}
+\frac{3}{r}\frac{{\rm d}\phi}{{\rm d}r} \ = \ {\cal V}^\prime \left(\phi,T\right) \,.
\end{equation}
For thermally-induced decay,
\begin{equation}
S_{3} \ = \ \int {\rm d}^{3}x
\left[ \left(\nabla\phi\right)^{2} + {\cal V} \left(\phi,T\right)\right] \,,
\label{action-s3}
\end{equation}
where $\phi(r)$ is the $O(3)$-symmetric solution of
\begin{equation}
\frac{{\rm d}^{2}\phi}{{\rm d}r^{2}}
+\frac{2}{r}\frac{{\rm d}\phi}{{\rm d}r} \ = \
{\cal V}^\prime \left(\phi,T\right) \,.
\end{equation}

\subsection{Gravitational waves}
The GW signal from a FOPT consists of three main components: the scalar field
contribution during the collision of bubble walls~\cite{Kosowsky:1991ua, Kosowsky:1992vn, Huber:2008hg, Kosowsky:1992rz, Kamionkowski:1993fg, Caprini:2007xq}, the sound wave in the plasma after bubble collisions~\cite{Hindmarsh:2013xza, Giblin:2013kea, Giblin:2014qia, Hindmarsh:2015qta}, and the MHD turbulence in the plasma after bubble collisions~\cite{Caprini:2006jb, Kahniashvili:2008pf, Kahniashvili:2008pe, Kahniashvili:2009mf, Caprini:2009yp}.
Assuming the three components can be linearly superposed, the
total strength of GWs produced reads
\begin{equation}
h^{2}\Omega_{\text{GW}} \ \simeq \ h^{2}\Omega_{\phi}+h^{2}\Omega_{\text{SW}}+h^{2}\Omega_{\text{MHD}} \,.
\label{GW-total}
\end{equation}
Note that the global $U(1)$ symmetry breaking could also generate cosmic
strings, which then annihilate to produce GWs~\cite{Cui:2017ufi, Cui:2018rwi}. However, this effect turns out to be subdominant for the energy scales under consideration
here.

The envelope approximation is often used to calculate the GWs from the scalar $\phi$ contribution, and numerical simulations tracing the envelope of thin-walled bubbles reveal that~\cite{Huber:2008hg, Caprini:2015zlo}\footnote{For an analytic derivation of GW production in the thin-wall and envelope approximation, see e.g. Refs.~\cite{Jinno:2016vai, Jinno:2017fby}.}
\begin{eqnarray}
\nonumber
h^{2}\Omega_{\phi}\left(f\right) & \ \simeq \ & 1.67\times 10^{-5}\left(\frac{H_{*}}{\beta}\right)^{2}\left(\frac{\kappa_\phi \alpha}{1+\alpha}\right)^{2}\left(\frac{100}{g_{*}}\right)^{1/3}
\left(\frac{0.11v_{w}^{3}}{0.42+v_{w}^{2}}\right) S_{\text{env}}\left(f\right),
\label{GW-scalar}
\end{eqnarray}
where $f$ is the frequency; $g_\ast$ is the number of relativistic degrees of freedom in the plasma at the temperature $T_\ast$ when the GWs are generated;
$H_\ast$ is the Hubble parameter at $T_\ast$; $v_w$ is the bubble wall velocity in the rest frame of the fluid; $\alpha \equiv {\rho_{\rm vac}}/{\rho^\ast_{\rm rad}}$ is the ratio of the vacuum energy density  $\rho_{\rm vac}$ released in the PT to that of the radiation bath $\rho^\ast_{\rm rad} = g_\ast \pi^2 T_\ast^4/30$;
$\beta/H_{*}$ measures the rate of the PT;
$\kappa_\phi$ measures the fraction of vacuum energy that is converted to
gradient energy of the $\phi$ field; and  $S_{\text{env}}(f)$ parameterizes the spectral shape of the GW radiation,
\begin{equation}
S_{\text{env}}\left(f\right) \ = \
\frac{3.8\left(f/f_{\text{env}}\right)^{2.8}}
{1+2.8\left(f/f_{\text{env}}\right)^{3.8}} \,.
\end{equation}
The peak frequency $f_{\rm env}$ of the $\phi$ contribution to the spectrum is determined by $\beta$ and by the peak frequency
$f_{\ast} = 0.62\beta /(1.8-0.1v_{w}+v_{w}^{2})$~\cite{Huber:2008hg} at the time of GW production,
\begin{equation}
f_{\text{env}} \ = \ \left(\frac{f_{*}}{\beta}\right)\left(\frac{\beta}{H_{*}}\right)h_{*} \,.
\end{equation}
Assuming the Universe is radiation-dominated after the PT and has expanded adiabatically ever since, the inverse Hubble time $h_\ast$ at GW production, red-shifted to today, is
\begin{equation}
h_{\ast} \ = \
16.5\times10^{-3} \text{mHz} \left(\frac{T_{*}}{100 \text{ GeV}}\right)
\left(\frac{g_{*}}{100}\right)^{1/6} \,.
\end{equation}

The SW contribution is given by~\cite{Hindmarsh:2015qta}:
\begin{eqnarray}
\nonumber
h^{2}\Omega_{\text{SW}}\left(f\right) & \ \simeq \ & 2.65\times10^{-6}\left(\frac{H_{*}}{\beta}\right)
\left(\frac{\kappa_{v}\alpha}{1+\alpha}\right)^{2}
\left(\frac{100}{g_{*}}\right)^{1/3}
v_{w}S_{\text{SW}}\left(f\right)
\label{GW-sw}
\end{eqnarray}
where $\kappa_v$ is the fraction of vacuum energy that is converted to bulk motion of the fluid, and the spectral shape
\begin{equation}
S_{\text{SW}}\left(f\right) \ = \ \left(\frac{f}{f_{\text{SW}}}\right)^{3}\left(\frac{7}{4+3\left(f/f_{\text{SW}}\right)^{2}}\right)^{7/2},
\end{equation}
with the peak frequency
\begin{equation}
f_{\text{SW}} \ = \
1.008\times \frac{2}{\sqrt{3}v_w}
\left(\frac{\beta}{H_{*}}\right)h_{*} \,.
\end{equation}

The MHD turbulence contribution is given by~\cite{Caprini:2009yp, Binetruy:2012ze}:
\begin{eqnarray}
\nonumber
h^{2}\Omega_{\text{MHD}}\left(f\right) & \ \simeq \ & 3.35\times10^{-4}\left(\frac{H_{*}}{\beta}\right)
\left(\frac{\kappa_{\text{MHD}}\alpha}{1+\alpha}\right)^{3/2}
\left(\frac{100}{g_{*}}\right)^{1/3}
 v_{w}S_{\text{MHD}}\left(f\right) \,,
 \label{GW-mhd}
\end{eqnarray}
where $\kappa_{\rm MHD}$ is the fraction of vacuum energy that is transformed into MHD turbulence, and the spectral shape can be found analytically:
\begin{equation}
S_{\text{MHD}}\left(f\right) \ = \ \frac{\left(f/f_{\text{MHD}}\right)^{3}}
{\left[1+\left(f/f_{\text{MHD}}\right)\right]^{11/3}\left(1+8\pi f/h_{*}\right)},
\end{equation}
where the peak frequency measured today is
\begin{equation}
f_{\text{MHD}} \ = \ 0.935 \left( \frac{3.5}{2v_w} \right)
\left(\frac{\beta}{H_{*}}\right)h_{\ast} \,.
\end{equation}

In most of the parameter space of interest, the phase transition occurs in
the `runaway bubbles in the plasma' regime, where
the $\Omega_\phi$ contribution cannot be neglected. In principle, friction
from the plasma could stop the bubble wall at some terminal
velocity~\cite{Bodeker:2017cim}, thus rendering the plasma-related GW
contributions and their associated uncertainties more
important~\cite{Ellis:2018mja, Ellis:2019oqb}. The friction term, however,
turns out to be unimportant in our case, because the $\phi$ field only
couples to the Higgs doublet through a scalar quartic interaction, and does
not directly couple to the gauge fields. Therefore, the friction from the
plasma does not grow with energy~\cite{Bodeker:2017cim}, thus preserving the
runaway behavior. In this limit, the uncertainties in the SW and MHD
contributions to $\Omega_{\rm GW}$ due to nonlinearities developing in the
plasma do not significantly affect our results.

To calculate the GW signal $\Omega_{\rm GW}$ described above, we need to know
the following quantities:
\begin{itemize}
  \item The ratio $\alpha$ of vacuum energy density released in the PT to that of the radiation bath.

  \item The rate of the PT, $\beta / H_{*}$. The smaller $\beta / H_{*}$, the stronger the PT.
 From the bubble nucleation rate $\Gamma (t) = A(t) e^{-S_E(t)}$~\cite{Turner:1992tz}, with $A(t)$ the amplitude and $S_E$ the Euclidean action of a critical bubble, we have:
  \begin{equation}
  \label{eqn:beta}
  \beta \ \equiv \
  - \left. \frac{{\rm d}S_E}{{\rm d}t} \right|_{t=t_\ast} \ = \
  T H_\ast \left. \frac{{\rm d}S_E}{{\rm d}T} \right|_{T=T_\ast} \,,
  \end{equation}
  where we have assumed the nucleation temperature $T_n \simeq T_\ast$ (or equivalently $t_n \simeq t_\ast$ with $t_n$ and $t_\ast$ respectively the time for bubble nucleation and GW production).
  \item The latent heat fractions $\kappa$ for each of the three processes. For the case of runaway bubbles in a plasma, we have
      \begin{eqnarray}
	  \kappa_{\phi} \ = \ \frac{\alpha-\alpha_{\infty}}{\alpha} \,, \quad
	  \kappa_v \ = \ \frac{\alpha_\infty}{\alpha}\kappa_\infty \,, \quad
	  \kappa_{\text{MHD}} \ = \ \epsilon \kappa_v \,,
	  \end{eqnarray}
      where $\epsilon$ is the turbulent fraction of bulk motion, which is
		found to be at most ($5\% - 10\%$)~\cite{Hindmarsh:2015qta}. To be concrete, we choose $\epsilon=0.1$ in this paper. We also have:
      \begin{eqnarray}
	  \kappa_{\infty} & \ \equiv \ &
      \frac{\alpha_{\infty}}{0.73+0.083\sqrt{\alpha_{\infty}}
      +\alpha_{\infty}} \,, \\
	{\rm with}~  \alpha_{\infty} & \ \simeq \ &
      \frac{30}{24\pi^{2}}\frac{\sum_{i} c_{i} \Delta m_{i}^{2}}{g_{*}T_{*}^{2}} \,.
      \label{alpha-infty}
	  \end{eqnarray}
      In Eq.~(\ref{alpha-infty}), the sum runs over all particles $i$ that are light in the initial phase and heavy in the final phase; $\Delta m_{i}^{2}$ is the squared mass difference in the two phases; and $c_i = n_i \, (n_i/2)$ for bosons (fermions) with $n_i$ the number of degrees of freedom of the
particle~\cite{Espinosa:2010hh}.
		
	\item The bubble wall velocity $v_w$ in the rest frame of the fluid away from the bubble. A conservative estimate for $v_w$ is given by~\cite{Steinhardt:1981ct}:
	\begin{equation}
	v_{w}=\frac{1/\sqrt{3}+\sqrt{\alpha^{2}+2\alpha/3}}{1+\alpha}.
	\end{equation}
	
	\item The number of relativistic degrees of freedom $g_*$ at the time of the PT, which is taken to be the SM contribution of 106.75 plus an additional 1 from the ALP.
\end{itemize}

\subsection{Critical temperature}
\label{subsec:critical-temp}
For the calculation of the GW production associated to the PT, we take $f_a$,
$\kappa$ and $\lambda_a$ as the only free parameters in the scalar potential (\ref{eqn:tree-level-V}).
Below the scale $f_a$, the scalar sector only contains the SM Higgs and the
superlight ALP $a$. The value of $\lambda_a$, as well as the other relevant coupling constants appearing in Eqs. (\ref{thermal-mass-h}) and (\ref{thermal-mass-phi}), at a high energy scale {$\Lambda < f_a$} can be obtained by running the SM RG equations up to the scale
$\Lambda$~\cite{Arason:1991ic}.  We will consider values $f_a \leq 10^8$ GeV, as the SM vacuum becomes unstable for $\Lambda \gtrsim 10^8$ GeV~\cite{Degrassi:2012ry} with the current best-fit top quark mass
$m_t = 173.0$ GeV~\cite{Tanabashi:2018oca}.  Similarly, we
take $f_a \geq 10^3$ GeV, because for $f_a$ comparable to (or smaller than) the Higgs mass, the LHC Higgs data impose stringent constraints on the coupling $\kappa$ (see  Section~\ref{sec:collider}).

\begin{figure*}
  \begin{centering}
  \includegraphics[height=0.23\textheight]{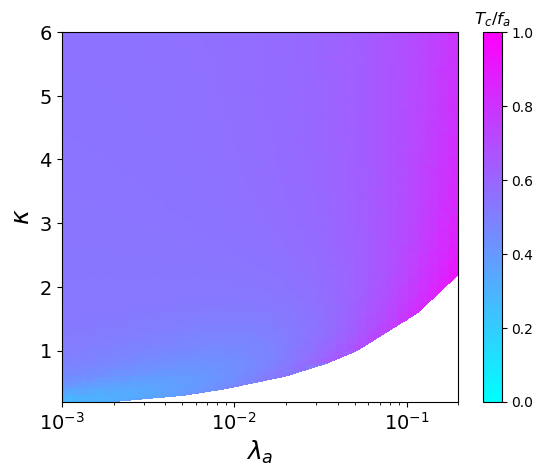}
  \includegraphics[height=0.23\textheight]{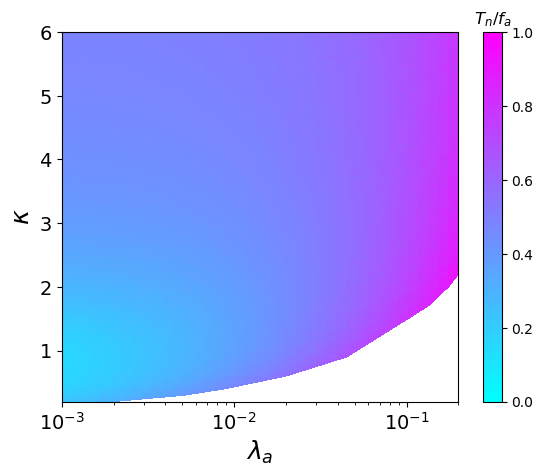}
  \par\end{centering}
  \caption{\label{fig:param-space-Tc} Critical temperature $T_c$ (left) and nucleation temperature $T_n$ (right) in units of $f_a$ in $(\kappa, \lambda_a)$ 	parameter space for  $f_a=10^6$ GeV.}
\end{figure*}

Fixing $f_a = 10^6$ GeV, we scan the two parameters $\kappa$ and $\lambda_a$.
The critical temperature is shown in the left panel of
Fig.~\ref{fig:param-space-Tc}, in units of $f_a$.
It can be seen that $T_c$ gets larger for larger $\kappa$ or larger $\lambda_a$, and in the parameter region we focus on, we have $T_c \lesssim f_a$.

\subsection{Bubble nucleation}
\label{subsec:rate-nucleation}

For the region in parameter space where there is a valid $T_c$, bubble nucleation might occur when $T<T_c$, i.e., when the two local minima become non-degenerate. The nucleation temperature $T_n$ is estimated by the condition:
\begin{equation}
\frac{\Gamma(T_n)}{H(T_n)^4} \ = \ 1 \,,
\label{nucl-temperature}
\end{equation}
where the Hubble constant is given by \cite{Ellis:2018mja}:
\begin{equation}
H(T) \ = \
\frac{\pi T^2}{3M_{\text{Pl}}}\sqrt{\frac{g_*}{10}} \,.
\end{equation}
If a valid solution for $T_n$ from Eq. (\ref{nucl-temperature}) can be found,
this indicates that the nucleation process will happen, and that the majority
of the GW signal is produced at $T_* \simeq T_n$. $\beta/H_*$
is obtained from Eq. (\ref{eqn:beta}), and the result for the
nucleation temperature $T_n$ is shown in the right panel of
Fig.~\ref{fig:param-space-Tc}, for the specific value $f_a = 10^6$ GeV.
It is clear from Fig.~\ref{fig:param-space-Tc} that the region with a viable
nucleation temperature $T_n$ is a subset of the FOPT region, and that
$T_n \lesssim T_c$.

\begin{figure*}[!t]
	\begin{centering}
		\includegraphics[width=0.44\textwidth]{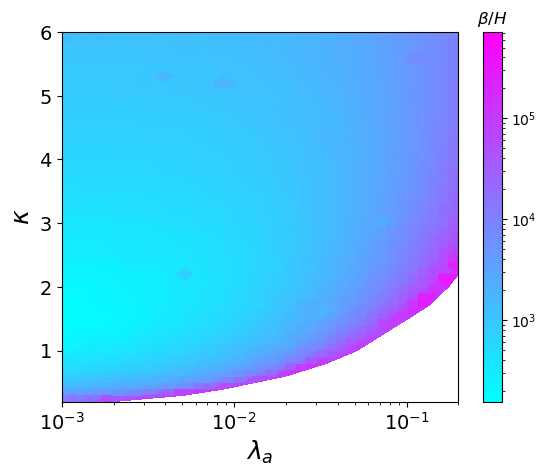}
		\includegraphics[width=0.44\textwidth]{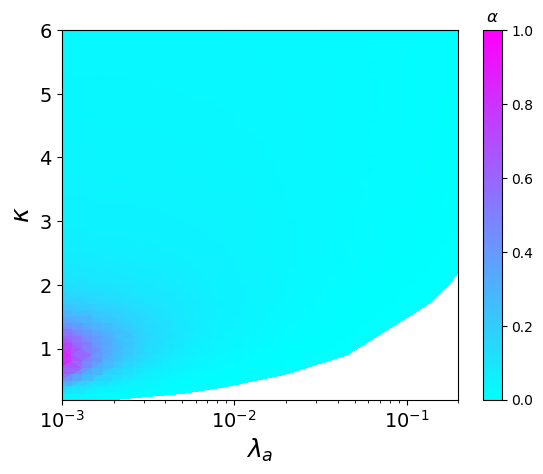}
		\par\end{centering}
	\caption{\label{fig:betaH-and-alpha} $\beta/H_*$ (left) and $\alpha$ (right) evaluated at $T_*$ for $f_a = 10^6$ GeV.}
\end{figure*}

The two important parameters for computing the GW signal, $\beta/H_*$ and $\alpha$, are then evaluated at $T_n$. We show the results of these two parameters in the parameter space $(\kappa, \lambda_a)$ for the case of $f_a =10^6$ GeV in Fig.~\ref{fig:betaH-and-alpha}. $\beta/H_*$ can reach values as small as
$\sim 10^2$; while $\alpha \lesssim 0.05$ for most of the parameter space,
but it can reach values of order 1. The region with large $\alpha$ and relatively small $\beta/H_*$ is where a relatively large GW signal is expected. This is the region where $0.001\lesssim \lambda_a \lesssim 0.1$ and $\kappa \sim {\cal O}(1)$. Note that larger values of $\kappa$ would lead to a breakdown of the perturbation theory, as a Landau pole is developed below the $f_a$ scale. Moreover, the one-loop self-energy corrections to the $\phi$ field due to the Higgs and Goldstone modes will be large in this case. In order to avoid these theoretical issues, we will restrict ourselves to $\kappa < 6$.

\subsection{Detection prospects}

We have assumed that the majority of the GW signal is produced at
$T_\ast \simeq T_n$.
Since the three GW contributions scale as inverse powers of $\beta/H_\ast$,
\begin{eqnarray}
h^2 \Omega_{\phi} \ \propto \ \left(\frac{\beta}{H_*}\right)^{-2} \,, \;\;
h^2 \Omega_{\rm SW} \ \propto \  \left(\frac{\beta}{H_*}\right)^{-1} \,, \;\;
h^2 \Omega_{\rm MHD} \ \propto \  \left(\frac{\beta}{H_*}\right)^{-1} \,,
\end{eqnarray}
we expect that a relatively large GW signal can be generated in
the small-$\beta/H_*$ region.

\begin{figure*}
  \begin{centering}	\includegraphics[width=0.485\textwidth]{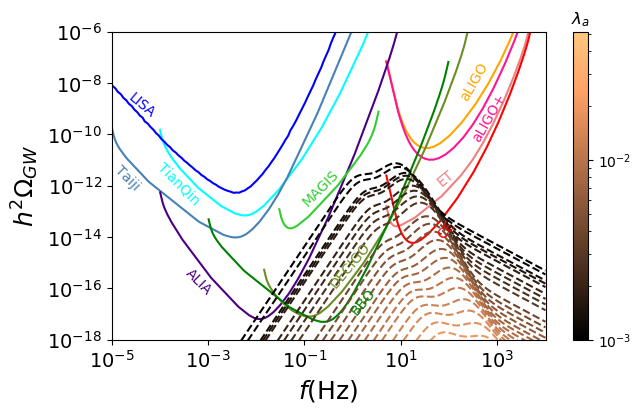} \\
  \includegraphics[width=0.485\textwidth]{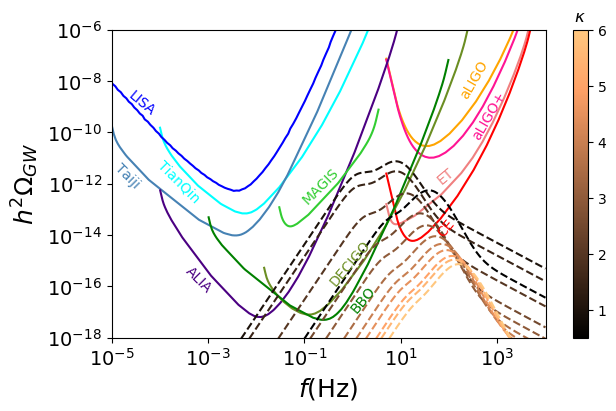}
  \includegraphics[width=0.485\textwidth]{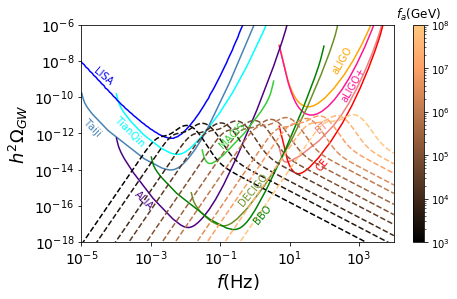}
  \par\end{centering}
  \caption{\label{fig:signal-vary}
  The detection prospects for the GW experiments
  \TianQin, \Taiji, \LISA, \ALIA,
  \MAGIS, \DECIGO, \BBO,
  \aLIGO, \aLIGOp, \ET ~and \CE, and the curves of GW strength $h^2 \Omega_{\text{GW}}(f)$ as functions of the three parameters $f_a$, $\kappa$ and $\lambda_a$ in the ALP model. In the upper panel, we have fixed $f_a=10^6$ GeV and $\kappa=1.0$ and varied $\lambda_a$ from $0.001$ to $0.2$; in the lower left panel $f_a=10^6$ GeV and $\lambda_a=0.001$, with $\kappa$ varying from $1.0$ to $6.00$; in the lower right panel $\kappa=1.0$ and $\lambda_a=0.001$, with $f_a$ between $10^3$ GeV and $10^8$ GeV. }
\end{figure*}

The three different components of GW signals from bubble wall collision in Eq.~\eqref{GW-scalar}, SW in the plasma in Eq.~\eqref{GW-sw}, and MHD turbulence in Eq.~\eqref{GW-mhd} are added to obtain the total GW emission, given in Eq.~\eqref{GW-total}, as a function of frequency $f$. 
Our numerical simulations reveal that the GW signal obtained in this model can be as large as $h^2 \Omega_{\text{GW}} \approx 10^{-12}$ for configurations with $\kappa \approx 1$ and $\lambda_a \approx 10^{-3}$. For such configurations that can produce considerable GW signals, contributions from both scalar fields and sound waves are comparable, and cannot be neglected, as indicated by the shape of the $h^2 \Omega_{\text{GW}}$ curves in Fig.~\ref{fig:signal-vary}. In most cases we have $h^2\Omega_{\text{SW}} \gtrsim h^2\Omega_{\phi}$ and the MHD contribution is much smaller than the other two contributions.

In Fig.~\ref{fig:signal-vary}, we show the detection prospects for the
future GW experiments \TianQin, \Taiji, \LISA, \ALIA,
\MAGIS, \DECIGO, \BBO,
\aLIGO, \aLIGOp, \ET ~and \CE.
To see the dependence of the GW signal on the parameters $f_a$, $\kappa$ and $\lambda_a$, let us first fix $f_a = 10^6$ GeV and $\kappa = 1.0$ and vary the quartic coupling $\lambda_a$ from $0.001$ to $0.2$. The corresponding GW signal $h^2 \Omega_{\rm GW}$ is shown in the upper panel of Fig.~\ref{fig:signal-vary}, as a function of the frequency $f$. It is obvious that within the $T_*$ region the configurations with smaller $\lambda_a$ tend to produce a larger GW signal with a relatively smaller peak frequency, which is preferred by the space-based experiments.
Likewise, when we fix $f_a=10^6$ GeV and $\lambda_a=0.001$, the configurations with $\kappa \approx 1$ tend to produce the strongest GW signals at small peak frequencies, as shown in the lower left panel of Fig.~\ref{fig:signal-vary}. When $\kappa$ and $\lambda_a$ are fixed, e.g. $\kappa=1.00$ and $\lambda_a=0.001$, a larger $f_a$ tends to produce a slightly larger GW signal with a larger peak frequency, as seen in the lower right panel of Fig.~\ref{fig:signal-vary}.

\begin{figure*}[!t]
  \begin{centering}
  \includegraphics[width=0.43\textwidth]{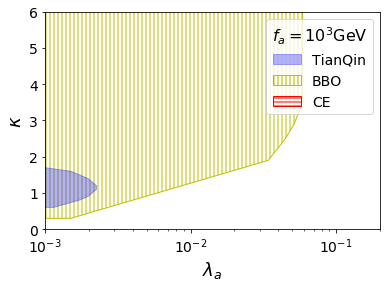}
  \includegraphics[width=0.43\textwidth]{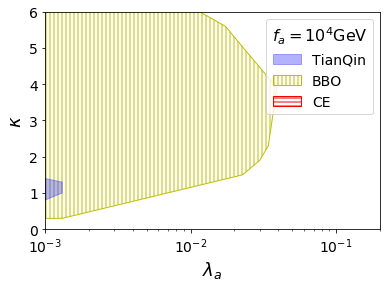}
  \includegraphics[width=0.43\textwidth]{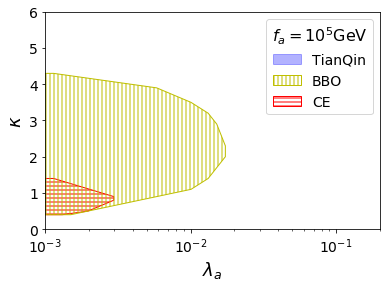}
  \includegraphics[width=0.43\textwidth]{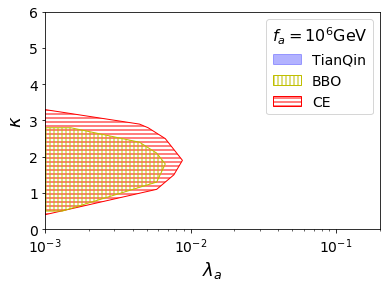}
  \includegraphics[width=0.43\textwidth]{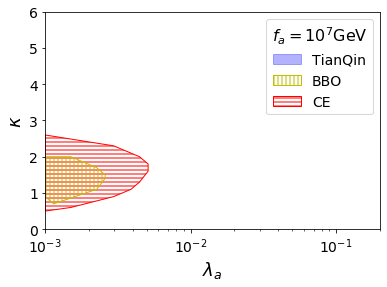}
  \includegraphics[width=0.43\textwidth]{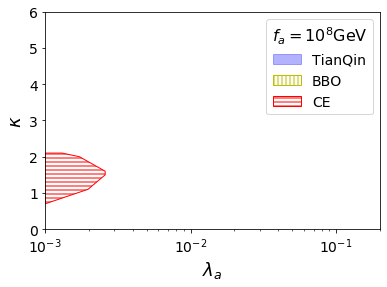}
  \par\end{centering}
  \caption{\label{fig:detect-region} GW detection prospects for
  \TianQin, \BBO ~and \CE ~in $(\kappa,\lambda_a)$ parameter space for $f_a = 10^{3,4,5,6,7,8}$ GeV. }
\end{figure*}

The GW detection prospects in the two-dimensional plane of $\kappa$ and
$\lambda_a$ are shown in Fig.~\ref{fig:detect-region} for the benchmark values $f_a = 10^{3,4,5,6,7,8}$ GeV. For the sake of clarity, we show only the sensitivity regions for three selected GW experiments: TianQin, BBO and CE.

\begin{figure}[!t]
  \begin{centering}
  \includegraphics[width=0.90\textwidth]{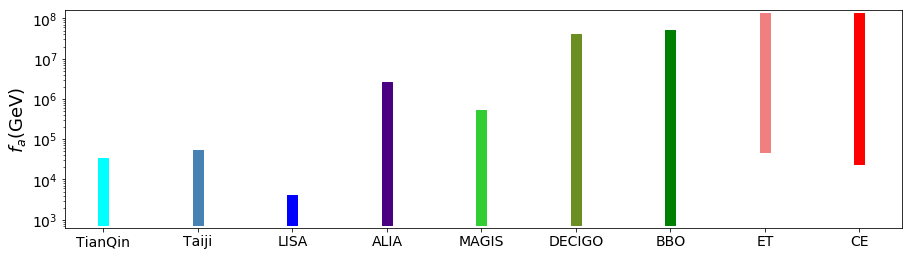}
  \par\end{centering}
  \caption{\label{fig:fa:range} Ranges of $f_a$ values accessible to the
	GW experiments
  	\TianQin, \Taiji, \LISA, \ALIA,
  	\MAGIS, \DECIGO, \BBO,
    \ET ~and \CE.
  We have fixed $\kappa=1.0$ and $\lambda_a=0.001$. The GW signal produced by this configuration is not accessible by \aLIGO ~or \aLIGOp.}
\end{figure}

The current and future GW observations are largely complementary to each
other. For instance, TianQin, Taiji, LISA ~and ALIA are more sensitive to the GWs with a comparatively lower frequency and thus a smaller $f_a$;  aLIGO, aLIGO+, ET and CE could probe higher frequency GWs, and thus larger
$f_a$; while MAGIS, DECIGO and BBO are able to cover the frequency range
in between. This is explicitly illustrated in Fig.~\ref{fig:fa:range}, for
the benchmark values $\kappa=1.0$ and $\lambda_a=0.001$. We use the power-law integrated sensitivity curves for future GW experiments, as described in
Appendix~\ref{app:A}. Although the GW emission in the ALP model does not
directly involve the ALP particle $a$, future GW observations could definitely probe a broad range of the decay constant $f_a$, which largely complements the low-energy, high-energy, astrophysical and cosmological constraints on the $f_a$ parameter, as detailed in Section~\ref{sec:complementarity}.
For related discussions on GW emission from the ALP field itself, see e.g.
Refs.~\cite{Machado:2018nqk, Ramberg:2019dgi, Croon:2019iuh}.

\section{Comparison with other ALP constraints}
\label{sec:complementarity}

As shown in Figs.~\ref{fig:signal-vary}, \ref{fig:detect-region} of Section~\ref{sec:GW}, current and future GW observations could probe a broad region of the parameter space in the ALP model. In particular, the scale $f_a$ in the range $(10^3 - 10^{8})$ GeV can be probed by future GW observatories, as summarized in Fig.~\ref{fig:fa:range}. At low energies, all the ALP couplings to SM particles are inversely proportional to powers of the decay constant $f_a$ (see e.g. Eqs.~(\ref{eqn:axion:couplings}) and (\ref{eqn:operator6})); thus, GW observations are largely complementary to the laboratory, astrophysical and cosmological constraints on the couplings of $a$ to SM particles. For the sake of simplicity, we will focus on the effective CP-conserving ALP couplings to photons ($g_{a\gamma\gamma}$), electrons ($g_{aee}$) and nucleons ($g_{aNN}$). In principle, the ALP could also couple to other SM particles like the muon, tau and other gauge bosons (gluons, $W$ and $Z$ boson), and we could even have CP-violating couplings to SM particles~\cite{Brivio:2017ije, Bauer:2017nlg, Bauer:2017ris} (see e.g. Refs.~\cite{Bauer:2017ris, Irastorza:2018dyq} for more details). In addition, the muon $g-2$ anomaly could be explained by ALP couplings to muons and photons~\cite{Bauer:2017nlg, Armillis:2008bg, deNiverville:2018hrc, Chiang:2018bnu} or by flavor violating couplings to muons and taus~\cite{Dev:2017ftk}.

\subsection{Low-energy effective ALP couplings}
\label{sec:axion:couplings}

Even though the ALP $a$ does not couple directly to the SM Higgs or the real scalar $\phi$ in the potential (\ref{eqn:tree-level-V}), low-energy couplings to SM particles can be induced at dimension-5 or higher. For instance, the effective couplings of $a$ to the SM photon and fermions $f$ can be written as
\begin{equation}
\label{eqn:axion:couplings}
{\cal L}_a \ = \
- \frac{C_{a\gamma}\alpha_{\rm EM}}{8\pi f_a} a F_{\mu\nu} \tilde{F}^{\mu\nu} + \frac{\partial_\mu a}{2f_a} \sum_f C_{af} (\bar{f} \gamma^\mu \gamma^5 f) \,.
\end{equation}
Here, $F_{\mu\nu}$ is the electromagnetic field strength tensor and $\tilde{F}^{\mu\nu}$ its dual, $\alpha_{\rm EM}$ is the fine-structure constant, and $C_{a\gamma}$, $C_{af}$ are model-dependent coefficients. Generally speaking, these coefficients are of order one for the QCD axion. Setting the model-dependent coefficients $C_i$ to one for simplicity, we rewrite the couplings in Eq.~(\ref{eqn:axion:couplings}) as
\begin{equation}
\label{eqn:axion:couplings2}
{\cal L}_a \ = \
- \frac{g_{a\gamma\gamma}}{4} a F_{\mu\nu} \tilde{F}_{\mu\nu} -
a \sum_f g_{aff} (i \bar{f} \gamma^5 f) \,,
\end{equation}
where the effective couplings are related to the high scale $f_a$ via
\begin{equation}
\label{eqn:axion:couplings3}
g_{a\gamma\gamma} \ = \ \frac{\alpha_{\rm EM}}{2\pi f_a} \,, \quad
g_{aff} \ = \ \frac{m_f}{f_a} \,,
\end{equation}
and $m_f$ is the corresponding fermion mass. We thus see that the GW limits on $f_a$ from Fig.~\ref{fig:fa:range} can be used to probe the effective couplings $g_{a\gamma\gamma}$, $g_{aee}$ and $g_{aNN}$.

\subsection{Coupling to photons}
\label{sec:gaa}

\begin{figure}[!t]
  \begin{centering}
  \includegraphics[width=0.49\textwidth]{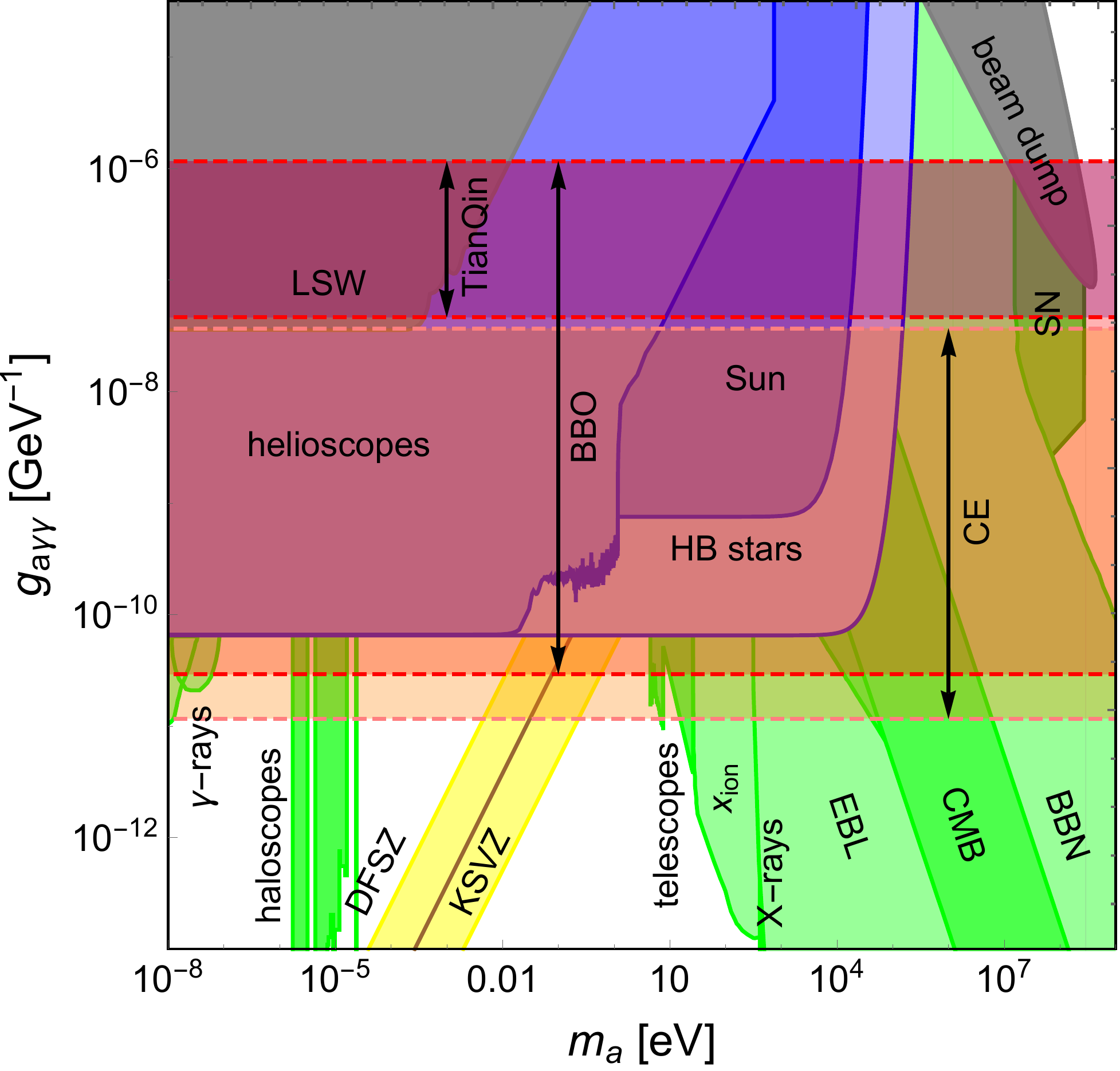}
  \includegraphics[width=0.49\textwidth]{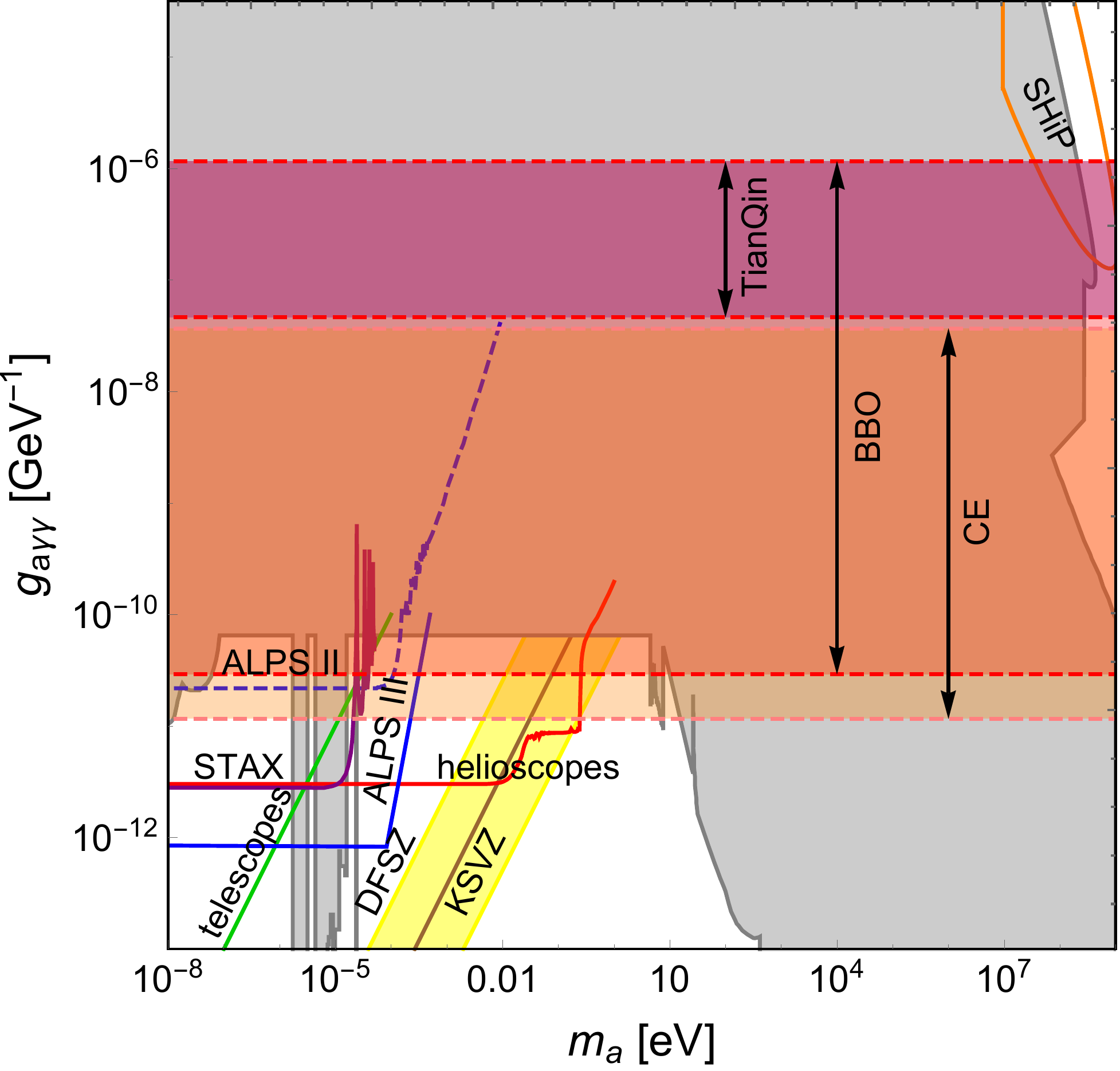}
  \par\end{centering}
  \caption{\label{fig:gaa} Complementarity between the GW limits on
	$g_{a\gamma\gamma}$ and laboratory, astrophysical and cosmological
	constraints on the ALP mass $m_a$ and $g_{a\gamma\gamma}$.
  The GW prospects for $g_{a\gamma\gamma}$ are shown in purple
	(\TianQin), red (\BBO) and orange (\CE), with dashed border lines.
  Other available constraints are collected in the left panel, including
	those from LSW experiments~\cite{Cameron:1993mr, Robilliard:2007bq, Chou:2007zzc, Afanasev:2008jt, Ehret:2010mh, Ballou:2015cka, Betz:2013dza},
	beam-dump experiments~\cite{Bergsma:1985qz, Bjorken:1988as, Riordan:1987aw, Blumlein:1990ay, Blumlein:1991xh}, helioscopes~\cite{Lazarus:1992ry, Moriyama:1998kd, Inoue:2002qy, Inoue:2008zp, Zioutas:2004hi, Andriamonje:2007ew, Arik:2008mq, Arik:2011rx, Arik:2013nya, Anastassopoulos:2017ftl}, observations of
	the Sun~\cite{Vinyoles:2015aba}, HB stars~\cite{Ayala:2014pea} and SN1987A (labelled as ``SN'')~\cite{Masso:1995tw}, telescope~\cite{Blout:2000uc, Hlozek:2014lca} and haloscope~\cite{DePanfilis:1987dk, Wuensch:1989sa, Hagmann:1990tj, Asztalos:2009yp} searches of ALP cold DM, and cosmological constraints
	from BBN, CMB, EBL, x-rays, $\gamma$-rays, $x_{\rm ion}$~\cite{Cadamuro:2010cz}. In the right panel, all the current limits are shown in gray, and
	we emphasize the future reach of telescope observations (green line)~\cite{Sigl:2017sew}, helioscope experiments (red line)~\cite{Anastassopoulos:2017kag, Irastorza:2011gs, Armengaud:2014gea}, the LSW experiments ALPS II (dashed blue line)~\cite{Bahre:2013ywa}, ALPS III (solid blue line)~\cite{ALPS-III}, STAX (solid purple line)~\cite{Capparelli:2015mxa} and SHiP (solid orange line)~\cite{Anelli:2015pba}. The regions above the lines can be probed by these
	experiments. In both the panels we also display the parameter space for DFSZ (yellow region) and KSVZ axions (brown line). The limits and prospects are adapted from~\cite{Irastorza:2018dyq} (see text for more details). }
\end{figure}

Following Ref.~\cite{Irastorza:2018dyq}, all the current constraints on the ALP couplings to photons $g_{a\gamma\gamma}$ are collected in the left panel of Fig.~\ref{fig:gaa}, while future laboratory and astrophysical prospects are shown in the right panel of Fig.~\ref{fig:gaa}. In both panels we also show the parameter space for DFSZ~\cite{Dine:1981rt, Zhitnitsky:1980tq} and KSVZ~\cite{Kim:1979if, Shifman:1979if} axions, indicated respectively by the yellow region and brown line. The various constraints are explained below:
\begin{itemize}
  \item Given the coupling $g_{a\gamma\gamma}$, the ALP can decay into two photons in the early universe, with a rate depending largely on its mass $m_a$ and the magnitude of $g_{a\gamma\gamma}$. If ALPs decay before recombination, the photons produced in the decays would potentially distort the cosmic microwave background (CMB) spectrum and, at earlier times, they would also affect big bang nucleosynthesis (BBN)~\cite{Cadamuro:2010cz}. The monochromatic photon lines from axion/ALP decays are also constrained by the flux of extragalactic background light (EBL) and direct searches in X-rays and $\gamma$-rays~\cite{Cadamuro:2010cz}. Furthermore, the photons might also change the evolution of the hydrogen ionisation fraction, $x_{\rm ion}$~\cite{Cadamuro:2010cz}. These limits are shown in greenish color in the left panel of Fig.~\ref{fig:gaa}.

  \item Assuming ALPs account for all the DM, the regions labelled as ``telescopes'' in the left panel of Fig.~\ref{fig:gaa} have been excluded by direct decaying DM searches in galaxies~\cite{Blout:2000uc, Hlozek:2014lca}. It is promising that future telescopes could probe couplings down to~\cite{Sigl:2017sew}
      \begin{equation}
      g_{a\gamma\gamma} \ \sim \
      \left( 10^{-12} \, {\rm GeV}^{-1} \right) \times \,
      \left( \frac{m_a}{10^{-6} \, {\rm eV}} \right)
      \left( \frac{d}{2 \, {\rm kpc}} \right)^{1/2} \;\;
      {\rm for} \;\;
      10^{-7} \, {\rm eV} \lesssim m_a \lesssim 10^{-4} \, {\rm eV} \,.
      \end{equation}
      Taking the distance to the ALP source to be $d \simeq 2\, {\rm kpc}$, the future sensitivity could reach $g_{a\gamma\gamma} \sim 10^{-13} \, {\rm GeV}^{-1}$. This is shown as the green line in the right panel of Fig.~\ref{fig:gaa}.

  \item As a result of the coupling $g_{a\gamma\gamma}$, ALPs can be produced and emitted copiously from dense stellar cores, thus affecting stellar evolution~\cite{Friedland:2012hj, Aoyama:2015asa, Dominguez:2017yhy}. Large portions of the $\left(m_a, g_{a\gamma\gamma}\right)$ parameter space have been excluded by measurements of the solar neutrino flux and helioseismology~\cite{Vinyoles:2015aba}, the ratio of horizontal branch (HB) to red giants in globular clusters~\cite{Ayala:2014pea}, and SN1987A neutrino data~\cite{Masso:1995tw}. These limits are labelled respectively as ``Sun'', ``HB stars'' and ``SN'' in the left panel of Fig.~\ref{fig:gaa}.

  \item In the presence of an electromagnetic field, the ALP can be converted to a photon through the $a\gamma\gamma$ coupling~\cite{Sikivie:1983ip}. The axion helioscope
		experiments Brookhaven~\cite{Lazarus:1992ry}, SUMICO~\cite{Moriyama:1998kd, Inoue:2002qy, Inoue:2008zp} and CAST~\cite{Zioutas:2004hi, Andriamonje:2007ew, Arik:2008mq, Arik:2011rx, Arik:2013nya, Anastassopoulos:2017ftl} aim to detect X-rays from $a - \gamma$ conversion in the Sun. The absence of a signal can be used to set the limits on $g_{a\gamma\gamma}$ labelled as ``helioscopes'' in the left panel of Fig.~\ref{fig:gaa} . Future experiments such as TASTE~\cite{Anastassopoulos:2017kag} and
		IAXO~\cite{Irastorza:2011gs, Armengaud:2014gea} could improve
		current constraints by over one order of magnitude,
        also shown by the solid red line in the right panel of Fig.~\ref{fig:gaa}.

  \item In a static magnetic field, ALP DM in the $\sim 10^{-6}$ eV
	  mass range can be converted into a microwave
	  photons~\cite{Sikivie:1983ip}. Narrow regions around this range have
	  been excluded by the ADMX
	  experiment~\cite{DePanfilis:1987dk, Wuensch:1989sa, Hagmann:1990tj, Asztalos:2009yp}. They are labelled as ``haloscopes'' in the left panel of
	  Fig.~\ref{fig:gaa}. Although we do not display these limits, let
	  us mention that future stages of ADMX could probe a very narrow
	  range around $m_a \sim 10 \ \mu{\rm eV}$~\cite{Rybka:2014xca}, while
	  the ABRACADABRA experiment might be sensitive to light ALPs with
	  mass
	  $10^{-14} \, {\rm eV} \lesssim m_a \lesssim 10^{-6} \, {\rm eV}$
	  and couplings down to
	  $g_{a\gamma\gamma} \sim 10^{-19} \, {\rm GeV}^{-1}$~\cite{Kahn:2016aff}.

  \item Light-shining-through-wall (LSW) experiments provide the most
	  stringent laboratory constraints on $g_{a\gamma\gamma}$ for a broad
	  range of ALP mass $m_a$. In such experiments,
	  ALPs can be produced from intense photon sources in the presence
	  of magnetic fields and then convert back into photons.
	  The LSW limits from BRFT~\cite{Cameron:1993mr}, BMV~\cite{Robilliard:2007bq}, GammaV~\cite{Chou:2007zzc}, LIPPS~\cite{Afanasev:2008jt}, ALPS~\cite{Ehret:2010mh}, OSQAR~\cite{Ballou:2015cka} and CROWS~\cite{Betz:2013dza} are collectively shown in the left panel of Fig.~\ref{fig:gaa}.
	  The LSW limits could be further improved by up to four orders of
	  magnitude by the experiments ALPS II~\cite{Bahre:2013ywa},
	  ALPS III~\cite{ALPS-III} and STAX~\cite{Capparelli:2015mxa}, as
	  indicated by the dashed blue, solid blue and solid purple
	  lines in the right panel of Fig.~\ref{fig:gaa}.
      There are also constraints from the polarization experiment
      PVLAS~\cite{Semertzidis:1990qc, Cameron:1993mr, DellaValle:2014xoa, DellaValle:2015xxa}
      and from fifth force searches~\cite{Hoskins:1985tn, Kapner:2006si, Decca:2007jq, Geraci:2008hb, Sushkov:2011zz, Smith:1999cr, Schlamminger:2007ht}, which are however weaker and thus not shown in Fig.~\ref{fig:gaa}.

  \item In beam-dump experiments, ALPs can be produced off photons. The
	  limits on $g_{a\gamma\gamma}$ from the experiments
	  CHARM~\cite{Bergsma:1985qz}, E137 \cite{Bjorken:1988as},
	  E141~\cite{Riordan:1987aw} and
	  NuCal~\cite{Blumlein:1990ay, Blumlein:1991xh} are comparatively
	  weaker than those from the astrophysical observations above,
	  excluding a region $g_{a\gamma\gamma} \gtrsim 10^{-7} \, {\rm GeV}^{-1}$
	  for ALP masses $m_a \sim ({\rm MeV} - {\rm GeV})$, as shown in
	  Fig.~\ref{fig:gaa}. The future experiment SHiP~\cite{Alekhin:2015byh, Anelli:2015pba} will extend the exclusion regions to higher $m_a$, but it
	  will not push to smaller couplings
	  $g_{a\gamma\gamma}$~\cite{Dobrich:2015jyk, Dobrich:2019dxc}, as
	  indicated in the right panel of Fig.~\ref{fig:gaa}.
	  The projected limit from NA62 is expected to be weaker and thus not
	  shown.
\end{itemize}

For sufficiently small $g_{a\gamma\gamma}$, the ALP might be long-lived and
decay outside the detectors in high-energy colliders.
There have been searches of single photon plus missing transverse energy $e^+ e^- \to \gamma + \slashed{E}_T$ at LEP~\cite{Abbiendi:2000hh, Heister:2002ut, Achard:2003tx, Abdallah:2003np} and $pp \to \gamma + \slashed{E}_T$ at LHC~\cite{Chatrchyan:2012tea, Aad:2012fw, CMS:2014mea, Aad:2014tda}. Similarly, observations of radiative decays of Upsilon mesons
$\Upsilon \to \gamma + \slashed{E}_T$ at CLEO~\cite{Balest:1994ch} and BaBar~\cite{delAmoSanchez:2010ac} can be used to set limits on $g_{a\gamma\gamma}$.
If the ALP decays promptly in the detectors, then we have the three photon
signature $e^+ e^-,\, p\bar{p},\, pp \to \gamma + a \to \gamma\gamma\gamma$
at LEP~\cite{Acciarri:1994gb, Anashkin:1999da},
Tevatron~\cite{triphoton:CDF} and LHC~\cite{Aad:2015bua, Aaboud:2017lxm}. These limits could be improved by one to two orders of magnitude at future
colliders such as Belle II~\cite{Kou:2018nap}, ILC~\cite{Baer:2013cma}, FCC-ee~\cite{Gomez-Ceballos:2013zzn} and at later stages of the
LHC~\cite{Mimasu:2014nea, Jaeckel:2015jla}. Benefiting from the large proton
number in heavy ions, the photon-photon luminosity can be greatly enhanced in
heavy-ion collisions compared to proton-proton colliders, and the current LHC
bounds on $g_{a\gamma\gamma}$ can be improved by two orders of magnitude with
ultra-peripheral heavy-ion collisions~\cite{Knapen:2016moh}. However, even
at future colliders, the prospective limits are still too weak, at the level
of $g_{a\gamma\gamma} \gtrsim 10^{-5} \, {\rm GeV}^{-1}$~\cite{Mimasu:2014nea, Jaeckel:2015jla, Knapen:2016moh}, and hence not shown. Additional collider and flavor factory constraints on the coupling $g_{a\gamma\gamma}$ can be found e.g. in
Refs.~\cite{Brivio:2017ije, Mariotti:2017vtv, CidVidal:2018blh}.

The following effective operators appear at dimension 6 and
7~\cite{Brivio:2017ije, Bauer:2017nlg, Bauer:2017ris}:
\begin{equation}
\label{eqn:operator6}
{\cal L}_a \ \supset \
\frac{C_6}{f_a^2} (\partial_\mu a) (\partial^\mu a) (H^\dagger H) +
\frac{C_7}{f_a^3} (\partial^\mu a) (H^\dagger iD_\mu H) (H^\dagger H),
\end{equation}
with $C_{6,\,7}$ the dimensionless Wilson coefficients and $D_\mu$ the
covariant derivative for the SM Higgs doublet. These operators
generate ALP couplings $haa$ to the SM Higgs and $haZ$ to the $Z$ boson,
which induce exotic decays of the SM Higgs, i.e. $h \to aa$ and $h \to aZ$
with the ALPs further decaying into two photons
$a \to \gamma\gamma$~\cite{Bauer:2017nlg, Bauer:2017ris}. In principle,
we can set limits on $g_{a\gamma\gamma}$ from the searches for exotic decays
of the SM Higgs $h \to aa \to 4\gamma$ and $h \to aZ \to \gamma\gamma \ell^+ \ell^-$ (with $\ell = e,\,\mu$). However, these limits depend on the
coefficients $C_{6,7}$ in Eq.~(\ref{eqn:operator6}), and we do not include
them in Fig.~\ref{fig:gaa}.

Using Eq.~(\ref{eqn:axion:couplings3}), the GW bounds on $f_a$ from
TianQin, BBO and CE can be converted to limits on the effective coupling $g_{a\gamma\gamma}$ that do not depend on the ALP mass $m_a$,
and are shown as the purple, red and orange horizontal bands with dashed
border lines in Fig.~\ref{fig:gaa} respectively. When combined, these GW
observations are sensitive to the range
\begin{equation}
10^{-11} \, {\rm GeV}^{-1} \ \lesssim \
g_{a\gamma\gamma} \ \lesssim \
10^{-6} \, {\rm GeV}^{-1} \,.
\end{equation}
As shown in Fig.~\ref{fig:gaa}, some of the regions within this range of $g_{a\gamma\gamma}$ have been excluded by astrophysical and cosmological observations and laboratory experiments, while some are still unconstrained.
Assuming that a GW signal from the PT in the ALP model is found in the near
future, then we would expect
a {\rm positive} signal in future ALP searches. This can be seen in the
right panel of Fig.~\ref{fig:gaa}
by the overlap between the regions covered by GW searches and those covered by telescopes~\cite{Sigl:2017sew}, helioscopes~\cite{Anastassopoulos:2017kag, Irastorza:2011gs, Armengaud:2014gea}, the LSW experiments ALPS II~\cite{Bahre:2013ywa},
ALPS III~\cite{ALPS-III} and STAX~\cite{Capparelli:2015mxa},
and beam-dump experiments like SHiP~\cite{Anelli:2015pba}.

\subsection{Coupling to electrons}
\label{sec:gae}

\begin{figure}[!t]
  \begin{centering}
  \includegraphics[width=0.6\textwidth]{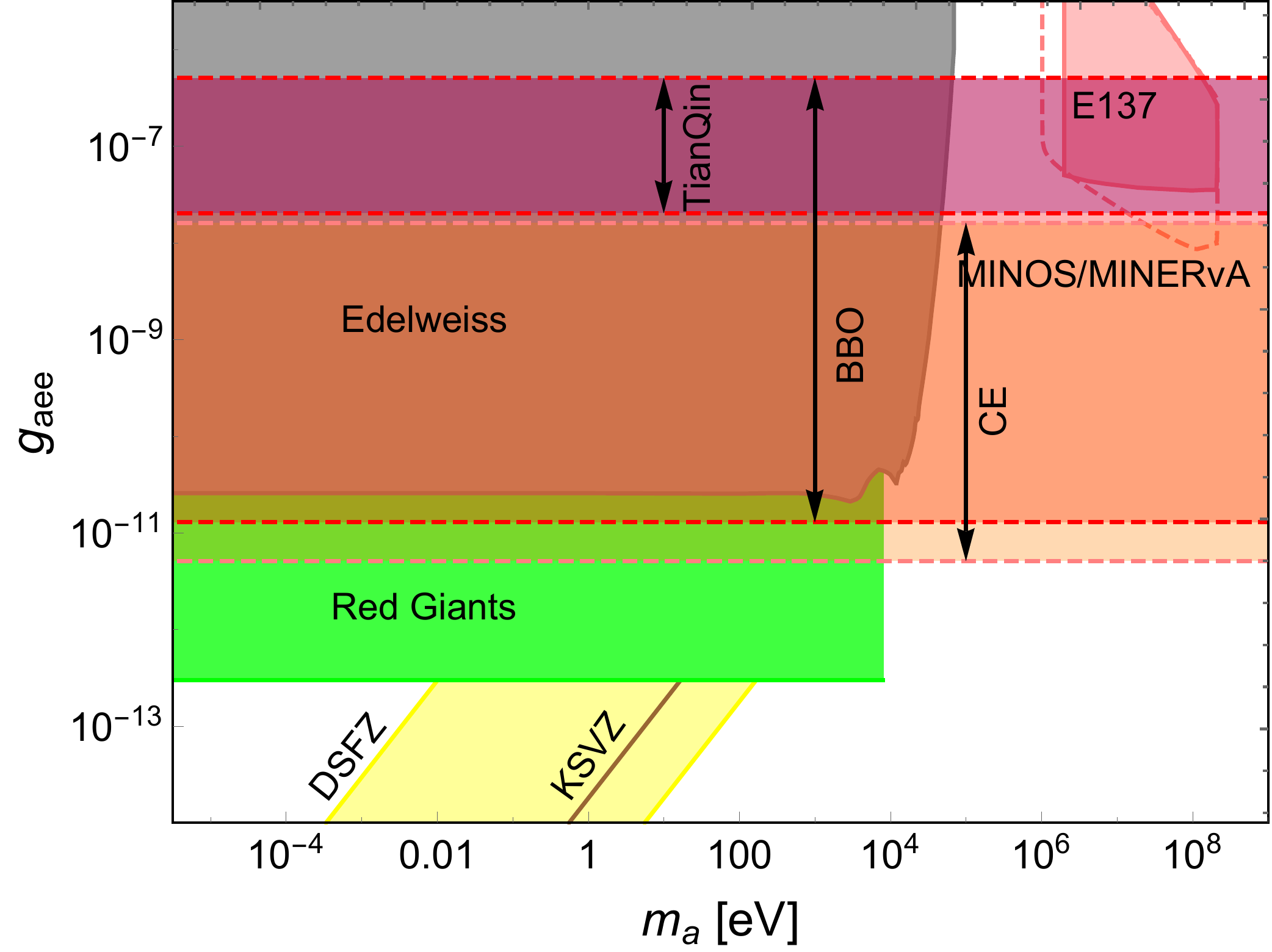}
  \par\end{centering}
  \caption{\label{fig:gae}
  Complementarity between the GW limits on $g_{aee}$ and laboratory and
	astrophysical constraints on the ALP mass $m_a$ and $g_{aee}$. The
	region of $g_{aee}$ values that will be probed by GW observatories
	is shown with dashed border lines for
	the case of \TianQin\ (purple),
	\BBO\ (red) and \CE\ (orange). The
	constraints include those from EDELWEISS
	(gray)~\cite{Armengaud:2013rta}, Red Giants (green)~\cite{Raffelt:2006cw}, and the beam-dump experiment
	E137 (pink)~\cite{Bjorken:1988as}. The dashed gray line indicates the
	prospect at MINOS/MINERvA~\cite{Essig:2010gu}. We also show the parameter space for DFSZ (yellow region) and KSVZ axions (brown line). See text for more details.
  }
\end{figure}

Astrophysical and laboratory constraints on the ALP coupling to electrons
$g_{aee}$ are collected in Fig.~\ref{fig:gae}. In this figure we also  show
the parameter space for DFSZ and
KSVZ axions. As in the case of
$g_{a\gamma\gamma}$, exotic SM Higgs decays $h \to aa$ and $h \to aZ$ can not
be used to set robust limits on the coupling $g_{ae}$, since they also depend
on the coefficients $C_{6,7}$ in Eq.~(\ref{eqn:operator6}). The other
constraints are described below:
\begin{itemize}
  \item ALPs can be produced by bremsstrahlung and Compton effects in the Sun.
	  Model-independent constraints on the mass $m_a$ and coupling
		$g_{aee}$ have been imposed by electron recoil searches
		in the low-background experiments Derbin~\cite{Derbin:2012yk},
		XMASS~\cite{Abe:2012ut} and EDELWEISS~\cite{Armengaud:2013rta}. 		The limit from EDELWEISS is the most stringent one, and is
		shown as the gray region in Fig.~\ref{fig:gae}.
		There  are also constraints from CoGeNT~\cite{Aalseth:2008rx}
		and CDMS~\cite{Ahmed:2009ht} on ALPs DM in local galaxies,
		which exclude however a much narrower region of $g_{aee}$.
		The limits on $g_{aee}$ from CUORE~\cite{Alessandria:2012mt},
		Derbin~\cite{Derbin:2011gg} and Borexino~\cite{Bellini:2012kz}
		depend on the effective ALP coupling
		to nucleons $g_{aNN}^{\rm eff}$, and are not shown in
		Fig.~\ref{fig:gae}.

  \item If the ALP couples to electrons, it will lead to extra energy losses
	  in astrophysical objects. Constraints from observations of solar
		neutrinos~\cite{Gondolo:2008dd} and Red
		Giants~\cite{Raffelt:2006cw} have excluded a broad region
		in parameter space. The Red Giant excluded region is shown
		in green in Fig.~\ref{fig:gae},
		while the solar neutrino limits are comparatively much weaker.

  \item ALPs can be produced in beam-dump experiments by bremsstrahlung off an incident electron beam and decay back into electron-positron pairs in the
	  detector~\cite{Essig:2010gu}. The region excluded by
		E137~\cite{Bjorken:1988as} is shown in pink in
		Fig.~\ref{fig:gae}. The experiment MINOS/MINERvA could
		improve significantly the current limit~\cite{Essig:2010gu},
		as indicated by the dashed gray line.
\end{itemize}

Given the relation in Eq.~(\ref{eqn:axion:couplings3}), the GW experiments
TianQin, BBO and CE could probe the range
\begin{equation}
10^{-11.5} \ \lesssim \ g_{aee} \ \lesssim \ 10^{-6.5} \,.
\end{equation}
A sizable fraction of this range for with $m_a \lesssim 10$ keV has already
been excluded by
EDELWEISS~\cite{Armengaud:2013rta} and Red Giants \cite{Raffelt:2006cw}.
Should the GW experiments TianQin find a GW signal, then the
corresponding ALP mass would be expected to be heavier than roughly $10$ keV,
which might be tested by the MINOS/MINERvA experiment~\cite{Essig:2010gu}.

\subsection{Coupling to nucleons}
\label{sec:gan}

\begin{figure}[!t]
  \begin{centering}
  \includegraphics[width=0.6\textwidth]{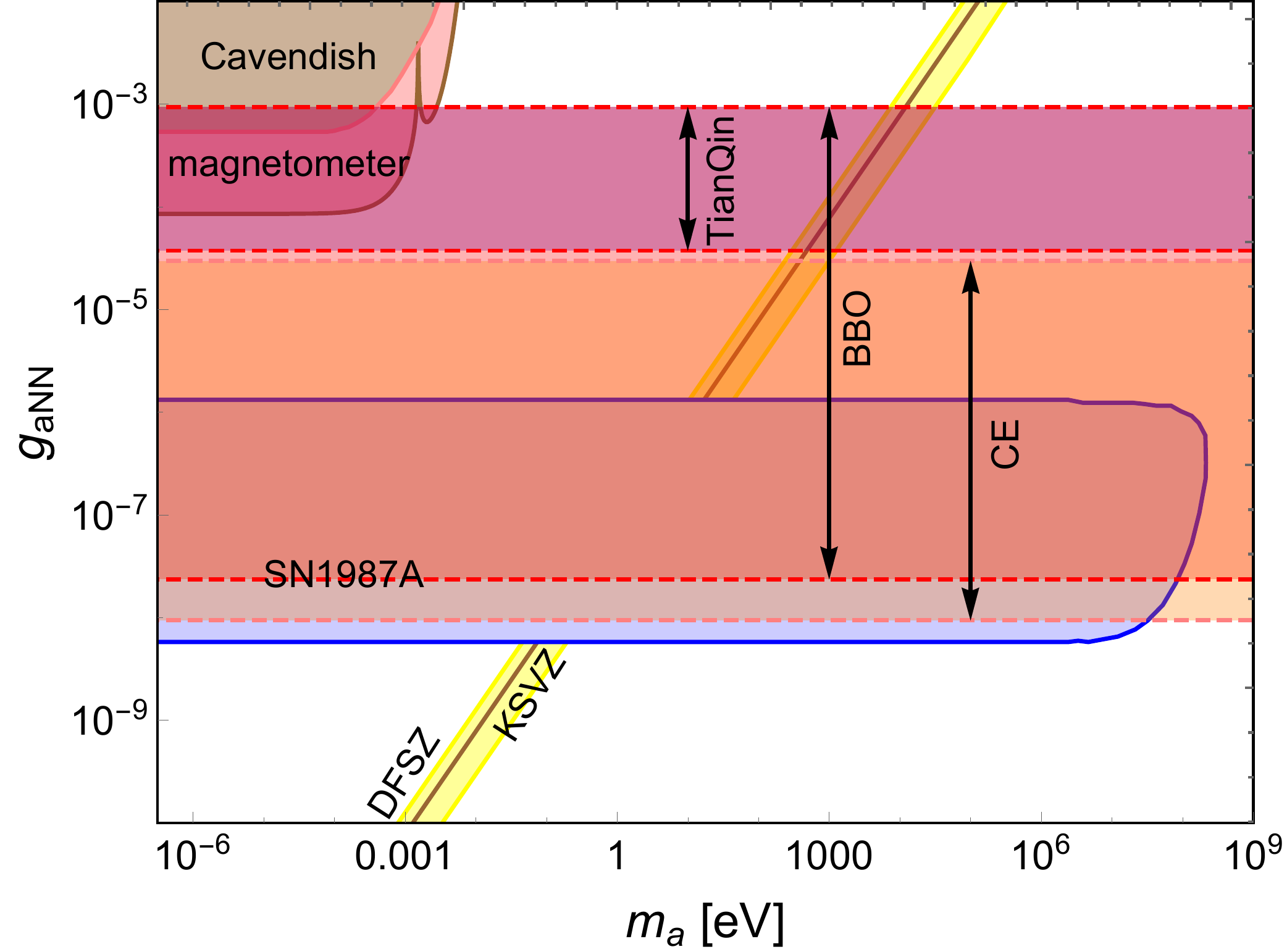}
  \par\end{centering}
  \caption{\label{fig:gan}
  Complementarity between GW limits on $g_{aNN}$ and astrophysical
	constraints on $m_a$ and $g_{aNN}$. GW experiments are sensitive
	to values of $g_{aNN}$ in the region shown in
	in purple (\TianQin), red (\BBO),
	 and orange (\CE), with dashed border lines.
	The constraints include those from
	SN1987A~\cite{Essig:2010gu} (blue), Cavendish-type
	experiments (brown) and magnetometer experiments (pink). We also
	show the parameter space for DFSZ (yellow region) and KSVZ axions (brown line). See text for more details.}
\end{figure}

Limits on the effective coupling $g_{aNN}$ of ALPs to nucleons are collected in Fig.~\ref{fig:gan}, which also displays the parameter space for DFSZ and KSVZ axions.
\begin{itemize}
  \item ALPs can be produced in compact astrophysical objects like neutron
	  stars and supernova cores via nucleon bremsstrahlung
		$N+ N \to N+ N + a$, where $N = p,\,n$ represents both protons
		and neutrons~\cite{Iwamoto:1984ir, Kaplan:1985dv}. Neutron
		star constraints on the  coupling $g_{aNN}$ can be found in
		e.g.~\cite{Iwamoto:1984ir, Sedrakian:2015krq}. Limits from
		neutrino bursts from SN1987A are stronger, and exclude
		couplings $10^{-8} \lesssim g_{aNN} \lesssim 10^{-6}$ for
		$m_a \lesssim 100$ MeV~\cite{Mayle:1987as, Burrows:1988ah, Raffelt:1987yt, Turner:1987by, Giannotti:2005tn, Chang:2018rso, Essig:2010gu}, as shown in blue in Fig.~\ref{fig:gan}. Next-generation supernova observations could improve greatly the limits on $g_{aNN}$,
		depending on how far the next supernova explosion
		is~\cite{Fischer:2016cyd}.
  \item The Yukawa couplings of ALPs to nucleons could potentially cause
	  violations of the gravitational inverse-square law, and the
		effective coupling $g_{aNN}$ is thus constrained by
		Cavendish-type
		experiments~\cite{Kapner:2006si, Adelberger:2006dh}, as shown
		in brown in Fig.~\ref{fig:gan}. Limits from measurements of
		Casimir forces are comparatively weaker with $g_{aNN} \lesssim 10^{-2.5}$~\cite{Klimchitskaya:2015zpa, Klimchitskaya:2017dsh}, and are not
		shown.

  \item Searches of new long-range spin-dependent forces between nucleons can be used to set limits on the coupling $g_{aNN}$. Magnetometer experiments have
	  excluded couplings $g_{aNN} \gtrsim 10^{-4}$ for ALP masses
		$m_a \lesssim$ meV~\cite{Vasilakis:2008yn}, as shown in pink
		in Fig.~\ref{fig:gan}.
\end{itemize}
The GW experiments TianQin, BBO and CE could probe the range
\begin{equation}
10^{-8} \ \lesssim \ g_{aNN} \ \lesssim \ 10^{-3} \,,
\end{equation}
which is largely complementary to supernova and laboratory constraints.

\section{Prospects from precision Higgs data at future colliders}
\label{sec:collider}

\begin{figure}[!t]
  \centering
  \includegraphics[width=0.55\textwidth]{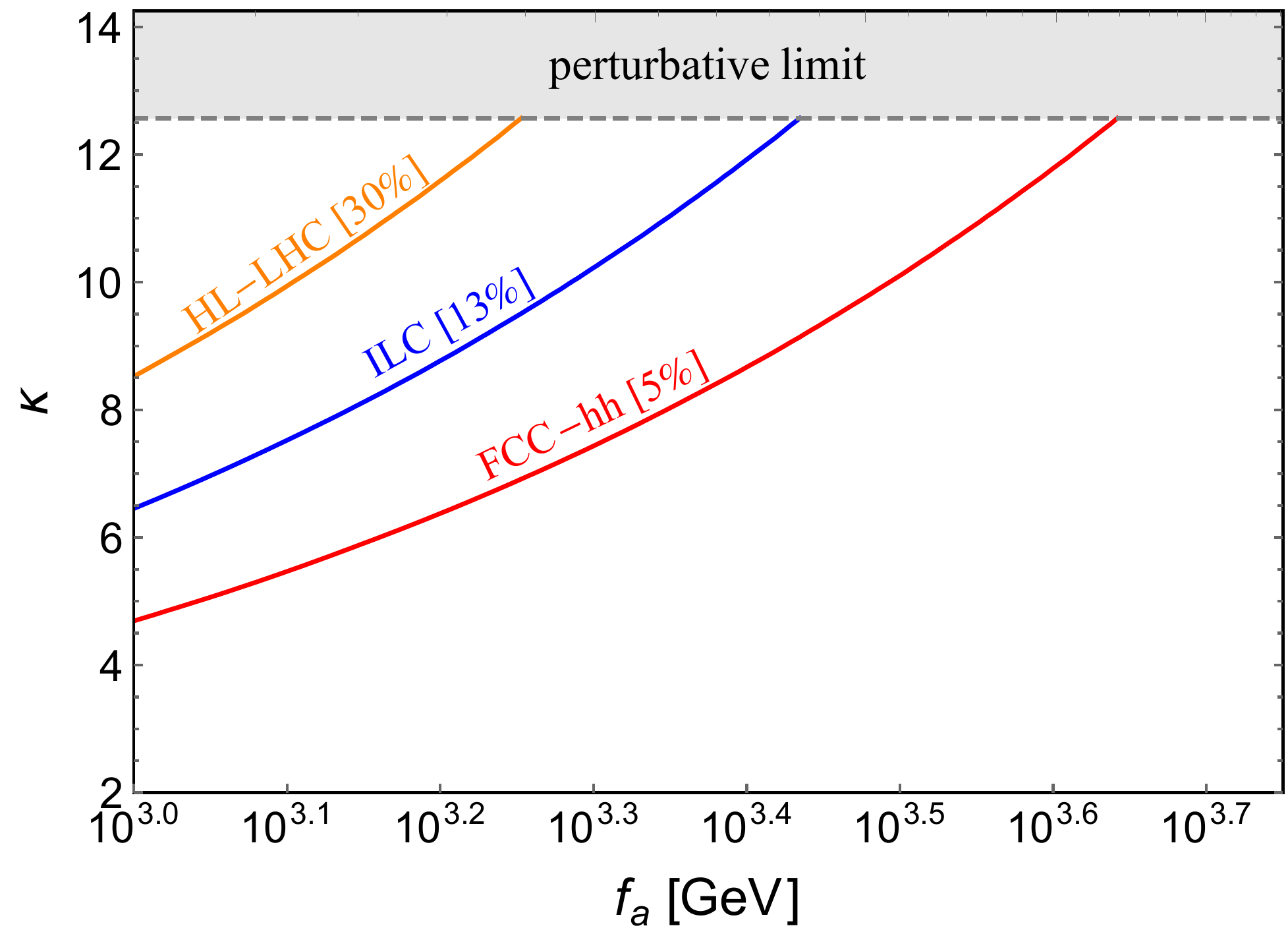}
  \caption{Prospects for the trilinear coupling of the SM Higgs at
	$1\sigma$ confidence level at the HL-LHC with center-of-mass energy
	$\sqrt{s} = 14$ TeV  and an integrated luminosity of 3
	ab$^{-1}$~\cite{Baglio:2012np, ATLAS2013, Goertz:2013kp, Yao:2013ika, Barger:2013jfa}, at the FCC-hh with $\sqrt{s} = 100$ TeV and
	30 ab$^{-1}$~\cite{Barr:2014sga}, and at the ILC with
	$\sqrt{s} = 1$ TeV and 2.5 ab$^{-1}$~\cite{Asner:2013psa, Tian:2013yda}, as a function of the scale $f_a$ and the coupling
	$\kappa$ in the ALP model. The shaded region is excluded by the
	perturbative limit. }
  \label{fig:trilinear}
\end{figure}

If the scalar $\phi$ in the ALP model resides at the few-TeV scale, it will contribute to the trilinear coupling $\lambda_3$ of the SM Higgs through the quartic coupling $\kappa$. This is obtained from the temperature-independent effective potential ${\cal V}_{}$ in Eq.~(\ref{eqn:potential:ft}), after integrating out the $\phi$-field, and reads~\cite{Curtin:2014jma, Huang:2015tdv}:
\begin{eqnarray}
\lambda_{3}  \ \simeq \
\lambda_3^{\rm SM} + \frac{\kappa^3 v_{\rm EW}^3}{24\pi^2 m_\phi^2} \,,
\end{eqnarray}
with the SM contribution $\lambda_3^{\rm SM} = m_h^2/2v_{\rm EW}$.
If the quartic coupling $\lambda_a \simeq {\cal O} (0.1) - {\cal O} (1)$ as
seen in Fig.~\ref{fig:detect-region}, the mass
$m_\phi = \sqrt{4\lambda_a} f_a$ is of the same order as the scale $f_a$. Then
we can set limits on the $f_a$ scale and $\kappa$ by precision measurements
of the trilinear SM Higgs coupling at high-energy colliders.
Current Higgs pair production data at the LHC lead to the limit
$-9 \lesssim \lambda_3 / \lambda_3^{\rm SM} \lesssim 15$~\cite{CMS:2017ihs, CMS:2017orf, Aaboud:2018knk}, which is too weak to exclude any parameter space
of the ALP model. Future hadron colliders like the high-luminosity LHC
(HL-LHC) and the FCC-hh~\cite{FCChh}, and lepton colliders such as the
ILC~\cite{Baer:2013cma}, will be able to measure the trilinear scalar coupling
more precisely and probe the scale $f_a$ and the quartic coupling $\kappa$.
Indeed, $\lambda_3$ can be measured within (30\% - 50\%) at the $1\sigma$
confidence level by the HL-LHC with an integrated luminosity of 3
ab$^{-1}$~\cite{Baglio:2012np, ATLAS2013, Goertz:2013kp, Yao:2013ika, Barger:2013jfa}. With a larger cross section, the precision can be improved to
$\sim 5$\% at the future 100 TeV collider FCC-hh with a luminosity of 30
ab$^{-1}$~\cite{Barr:2014sga}, and up to 13\% at the 1 TeV ILC with a luminosity of 2.5 ab$^{-1}$~\cite{Asner:2013psa, Tian:2013yda}.
All these sensitivities are shown in Fig.~\ref{fig:trilinear}, for the
benchmark value $\lambda_a = 0.25$. Future high energy colliders are largely
complementary to low energy axion experiments and GW observations for TeV
scale $f_a$.

\section{Conclusion}
\label{sec:conclusion}

In this paper we have studied the production of GWs due to a strong FOPT in a generic axion or ALP model, where we extended the SM scalar sector by adding only a complex singlet field $\Phi$. The angular component of $\Phi$ is identified as the axion or ALP field $a$. The original Lagrangian contains only a few free parameters, namely, the ALP mass $m_a$, the ``axion decay constant'' $f_a$ and the quartic couplings $\kappa$ and $\lambda_a$ in Eq.~(\ref{eqn:tree-level-V}). We have explored the prospects for GW emission for $f_a$ between $10^3$ GeV and $10^{8}$ GeV. Our numerical calculations reveal that in the ALP model we are considering, the GW signal strength could be as large as $h^2 \Omega_{\text{GW}} \sim 10^{-12}$, which might be detectable at future GW experiments like TianQin, BBO and CE, depending on the GW frequency and on the ALP model parameters (see Figs.~\ref{fig:signal-vary}-\ref{fig:fa:range}).

At low energies, the ALP couplings to SM particles are universally determined by the decay constant $f_a$, up to model-dependent coefficients; in other words, all the effective couplings of ALP depend on inverse powers of $f_a$. Therefore, we can convert the GW limits on $f_a$ to sensitivities on the effective ALP couplings to SM particles, independent of the ALP mass $m_a$. We have considered the CP-conserving couplings of ALP to photons $g_{a\gamma\gamma}$, electrons $g_{aee}$ and nucleons $g_{aNN}$. These couplings are tightly constrained by a large variety of laboratory experiments, and by astrophysical and cosmological observations, which exclude broad regions depending on the ALP mass $m_a$. GW experiments would probe sizable regions of the unconstrained parameter space, namely, $10^{-11} \, {\rm GeV}^{-1} \lesssim g_{a\gamma\gamma} \lesssim 10^{-6} \, {\rm GeV}^{-1}$, $10^{-11.5} \lesssim g_{aee} \lesssim 10^{-6.5}$ and $10^{-8} \lesssim g_{aNN} \lesssim 10^{-3}$, which are largely complementary to the laboratory, astrophysical and cosmological constraints. Thus, if a GW signal is found in future GW experiments and interpreted in the framework of axion or ALP models, it can be cross-checked in the upcoming laboratory and/or astrophysical ALP searches. In addition, for $f_a$ at the TeV scale, the real component $\phi$ contributes to the trilinear coupling of the SM Higgs. Thus precision Higgs data at future hadron and lepton colliders can be used to probe the $f_a$ and $\kappa$ parameters in the ALP model, which is also complementary to low-energy axion/ALP experiments and GW observations.

\section*{Acknowledgements}
We thank Robert Caldwell, Yanou Cui, Ryusuke Jinno, Arthur Kosowsky, Marek Lewicki, Andrew Long, Alex Pomarol, Michael Ramsey-Musolf and Fabrizio Rompineve for useful discussions and comments on the draft. B.D. also thanks Aniket Joglekar for a discussion on the trilinear Higgs coupling. This work was supported by the US Department of Energy under Grant No. DE-SC0017987. Y.C.Z. is grateful to the Center for High Energy Physics, Peking University where part of the work was done for generous hospitality.

\appendix
\section{Power-law Integrated Sensitivity Curves} \label{app:A}
Let us briefly outline the procedure used to compute the power-law integrated sensitivity curves for current and future GW experiments. For a detailed
description of this method, see Ref.~\cite{Thrane:2013oya}.

\begin{figure}[!t]
	\begin{centering}
		\includegraphics[width=0.6\textwidth]{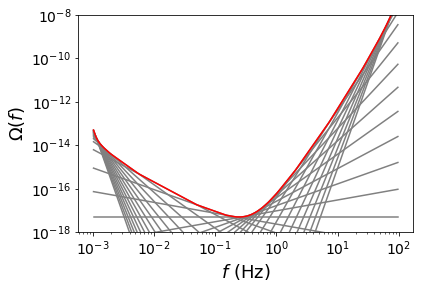}
		\par\end{centering}
	\caption{\label{fig:pi-curve} The sensitivity curves for BBO. The red curve is the power-law integrated sensitivity curve defined in Eq.~\eqref{pi-curve}; the gray curves are sensitivity curves for different power-law indices, defined in Eq.~\eqref{beta-curve}. }
\end{figure}

In the literature, the square root $\sqrt{S_n(f)}$ of the strain power spectral density is usually given as a function of frequency, in units of $1/\sqrt{\text{Hz}}$. We first convert it to $\Omega_n(f)$:
\begin{equation}
\Omega_n(f) \ = \ \frac{2\pi^2}{3H_0^2} f^3 S_n(f) \,.
\end{equation}
Then, given a set of power-law indices $\beta$, e.g. $\beta \in \{-8,-7,...,7,8\}$, we compute for each $\beta$
\begin{equation}
\Omega_{0\beta} \ = \ \frac{\rho}{\sqrt{2T}}
\left[
\int_{f_{\text{min}}}^{f_{\text{max}}} \text{d}f
\frac{(f/f_{\text{ref}})^{2\beta}}{\Omega_n^2(f)}
\right]^{-1/2} \,,
\end{equation}
where $f_{\text{ref}}$ is some reference frequency. It can be
arbitrarily chosen and it will not affect the results. $\rho$ is the integrated signal-to-noise ratio and $T$ is the observation time. Following
Ref.~\cite{Thrane:2013oya}, for $\rho=1$ and $T=1\text{ year}$ we have:
\begin{equation}
\label{beta-curve}
\Omega_\beta(f) \ = \ \Omega_{0 \beta}
\left(\frac{f}{f_{\text{ref}}} \right)^\beta \,.
\end{equation}
The power-law integrated sensitivity curve $\Omega_{\text{PI}}(f)$ is the envelope of all the $\Omega_\beta(f)$ curves,
\begin{equation}
\label{pi-curve}
\Omega_{\text{GW}}(f) \ = \ \max
\left[
\Omega_{0 \beta} \left(\frac{f}{f_{\text{ref}}} \right)^\beta
\right] \,.
\end{equation}
As an explicit example,  the power-law integrated curve $\Omega_{\text{GW}}(f)$ of BBO as well as the series of $\Omega_\beta(f)$ are presented in Fig.~\ref{fig:pi-curve}.

\end{document}